\documentclass[10pt]{article}

\usepackage[utf8]{inputenc} 
\usepackage[T1]{fontenc}

\usepackage{amsmath}
\usepackage{amsfonts}
\usepackage{amssymb}

\usepackage{lmodern}
\usepackage[a4paper]{geometry}
\usepackage{graphicx}
\usepackage{amsmath,amscd}
\usepackage{amssymb}
\usepackage{mathtools}
\usepackage{stmaryrd}
\usepackage{calrsfs}
\usepackage{pdfsync}
\usepackage[final]{pdfpages}
\usepackage{enumitem}
\usepackage{babel} 
\usepackage{esvect}
\usepackage{textcomp}
\usepackage{calc}
\usepackage{verbatim}
\usepackage[colorlinks=true,urlcolor=blue]{hyperref}
\usepackage{fancyhdr}
\usepackage{array}
\usepackage{appendix}
\usepackage{rotating}
\usepackage{multirow}
\usepackage{longtable}
\usepackage{caption} 
\usepackage[all]{xy}
\usepackage{romanbar}
\usepackage{listings}
\usepackage{color}

\def \NR{{\mathcal N}}

\def \v#1{\vspace{#1mm}}
\def \n{{\noindent}}

\def \1{{\bf 1}}

\def \begep{\begin{enumerate}[label=\arabic*)]}
\def \bege{\begin{enumerate}}
\def \ende{\end{enumerate}}

\def \blst{\begin{lstlisting}[language=R]}
\def \elst{\end{lstlisting}}

\begin{document}
\title{Note on a non-parametric method for change-point detection}
\author{Pierre Ailliot\footnote{Univ Brest, CNRS UMR 6205, Laboratoire de Mathématiques de Bretagne Atlantique, France.\newline
pierre.ailliot@univ-brest.fr
}\, ,
Jean-Marc Derrien\footnote{Univ Brest, CNRS UMR 6205, Laboratoire de Mathématiques de Bretagne Atlantique, France.\newline
jean-marc.derrien@univ-brest.fr
}\, ,
N'Dèye Coumba Niass\footnote{Univ Brest, CNRS UMR 6205, Laboratoire de Mathématiques de Bretagne Atlantique, France.}}
\date\today

\maketitle

\tableofcontents

\v4
\n\textbf{Keywords :} change-point, non-parametric test, Wilcoxon-Mann-Whitney, confidence interval, bootstrap

\section*{Abstract}

The purpose of this note is to present in details R codes to implement a non-parametric method for change-point detection. The proposed approach is validated from various perspectives using simulations. This method is a competitor to that of Pettitt (\cite{Pettitt}) and is, like the latter, based on the Wilcoxon-Mann-Whitney test. It is used in \cite{Valero} for the study of relatively short time series obtained from measurements on cores sampled in the bay of Brest. 

\newpage
\section{Introduction}

The statistical change-point detection method described here is based on repeated use of the Wilcoxon-Mann-Whitney (WMW) test.

The WMW test is a classical non-parametric test for comparing two independent samples in terms of stochastic dominance (all observations are assumed to be independent). As such, it can be used to assess the existence of a certain type of change in a time series at a pre-specified time point $\tau_0$: one sample consists of the first $\tau_0$ observations, and the other of the subsequent ones. The WMW test is distribution-free, meaning it does not require assumptions on the underlying distribution of the time series under the null hypothesis of no change, provided it is continuous.

In this note, we are interested in detecting a possible change in a time series without predefining a specific change-point $\tau_0$ — every time point $\tau$ is considered as a potential change-point. The test used to assess the presence of a change in a given time series is constructed as follows. Its statistic, denoted $V$, is defined as the minimum of the $p$-values obtained from WMW tests applied as described above to each possible change-point $\tau$ (except for the first and last five time points to avoid boundary effects). If this minimum is sufficiently close to zero, we interpret it as evidence of a change.

The distribution of $V$ under the null hypothesis $H_0$ (no change) is estimated via simulations. This allows us, given an observed value $v$ of $V$ for a particular time series, to estimate the probability that $V$ does not exceed $v$ under $H_0$ — this probability is the $p$-value of the test. The resulting test is, like the WMW test, distribution-free.

Its power, evaluated through simulations under various natural alternatives $H_1$, is found to be comparable to that of Pettitt’s non-parametric change-point test (\cite{Pettitt}), and very close to that of parametric tests restricted to their domain of validity.

When a change is detected by this test in a time series, the minimum $v$ of the WMW $p$-values occurs at a time point $\hat{\tau}$, which naturally serves as an estimate of the change-point.

To construct a confidence interval for this change-point, we use a non-parametric resampling approach as follows. Based on $\hat{\tau}$, we generate $10^3$ new time series by randomly resampling (with replacement) the first $\hat{\tau}$ values and the last $n - \hat{\tau}$ values of the original series (of length $n$), respectively.
For each of these $10^3$ series, we perform the same sequence of WMW tests and record the time point that minimizes the $p$-values. Simulations under $H_1$ show that the 0.025 and 0.975 quantiles of these $10^3$ estimated change-points provide an approximate 95\% confidence interval for the true change-point.

In the next section, we specify the statistical framework including the test assumptions (no change under $H_0$, presence of a change under $H_1$), and define the test statistic $V$. Section 3 is devoted to estimating the distribution of $V$ under $H_0$. In Section 4, we estimate the $p$-value and the possible change-point, and illustrate the method with an example. Section 5 discusses the power of the test, and the final section details the resampling method used to construct the confidence interval for the change-point.

R codes are discussed throughout this note. 

\section{Assumptions and test statistic for change-point detection}

The time series $(X_t)_{1\leq t \leq n}$ considered in this note are all of length $n = 57$, like most of the time series studied in \cite{Valero}.
The random variables $X_1, X_2, \ldots, X_n$ are assumed to be independent and continuously distributed.

\newpage
One of the goals of this note is to construct a statistical test for deciding between the null hypothesis:

{\leftskip=1.5cm
\n $H_0$: \flqq the random variables $X_1, X_2, \ldots, X_n$ are identically distributed\frqq;
\par}

\n and the alternative hypothesis:

{\leftskip=1.5cm
\n $H_1$: \flqq there exists a time point $\tau$, called the change-point, between $b$ and $n-b$ (with $b = 6$), such that the random variables $X_1, X_2, \ldots, X_\tau$ are identically distributed, as are $X_{\tau+1}, X_{\tau+2}, \ldots, X_n$, with the first group having a common distribution that differs from the common distribution of the second group\frqq.
\par}

\v3

To construct such a test, we begin by performing a sequence of $n - 2b + 1$ WMW tests comparing the two samples $X_1, X_2, \ldots, X_\tau$ and $X_{\tau+1}, X_{\tau+2}, \ldots, X_n$, for $\tau$ ranging from $b$ to $n - b$ (with $b = 6$, so that each $\tau$ allows for a potential $p$-value below 1~\textperthousand{}, according to the tables in \cite{MW}).

This results in a sequence of $n - 2b + 1$ $p$-values denoted as
$PVW\!MW_\tau$, for $b \leq \tau \leq n - b$.
The test statistic studied in this note is then defined as:
\begin{equation}
\label{eq:stattest}
V = \min_{b \leq \tau \leq n - b} PVW\!MW_\tau\, .
\end{equation}

The idea is to reject the null hypothesis $H_0$ when the observed value $v$ of $V$ is "too close" to 0.
An estimate of the distribution of $V$ under $H_0$ is obtained through simulations in the following section.

\section{Empirical distribution of the test statistic $V$ under $H_0$}

As indicated in the previous section,  throughout this note we perform numerical experiments
with time series of length $n=57$.  
The \lstinline|parallel| package is used repeatedly to parallelize computations in order to speed them up.  
The \lstinline|matrixTests| package will be used in the last section when it comes to determining  
a confidence interval for the estimation of the change-point by resampling.

\blst
n=57
library(parallel)
nclust = 4  # number of cores used (adjust according to your machine)
library(matrixTests)
\end{lstlisting}

We start by coding a first function \lstinline|pvWMW| which  
takes as input an integer \lstinline|k| and a time series \lstinline|x|  
and returns the $p$-value obtained by the WMW test applied  
to the two samples \lstinline|x[1:k]| and \lstinline|x[(k+1):length(x)]|.  Then
the function \lstinline|pvsWMW| also takes as input  
a time series \lstinline|x| and uses the function  
\lstinline|pvWMW| to return the sequence of $p$-values obtained  
by repeatedly applying the WMW test to the samples \lstinline|x[1:k]| and \lstinline|x[(k+1):length(x)]|  
for \lstinline|k| varying from \lstinline|b| to \lstinline|length(x)-b| with \lstinline|b=6|.  
Finally, the function {\ttfamily v} returns the value of the realization of the statistic $V$ defined by (\ref{eq:stattest}) corresponding to  
the time series \lstinline|x| given as input.
 
\blst
pvWMW=function(k,x){
    wilcox.test(x[1:k],x[(k+1):length(x)])$p.value
}

pvsWMW=function(x,b=6){
    sapply(b:(length(x)-b),pvWMW,x=x)
}

v=function(x){
    min(pvsWMW(x))
}
\end{lstlisting}

\begin{figure}[t]
    \centering
    \includegraphics[scale=0.5]{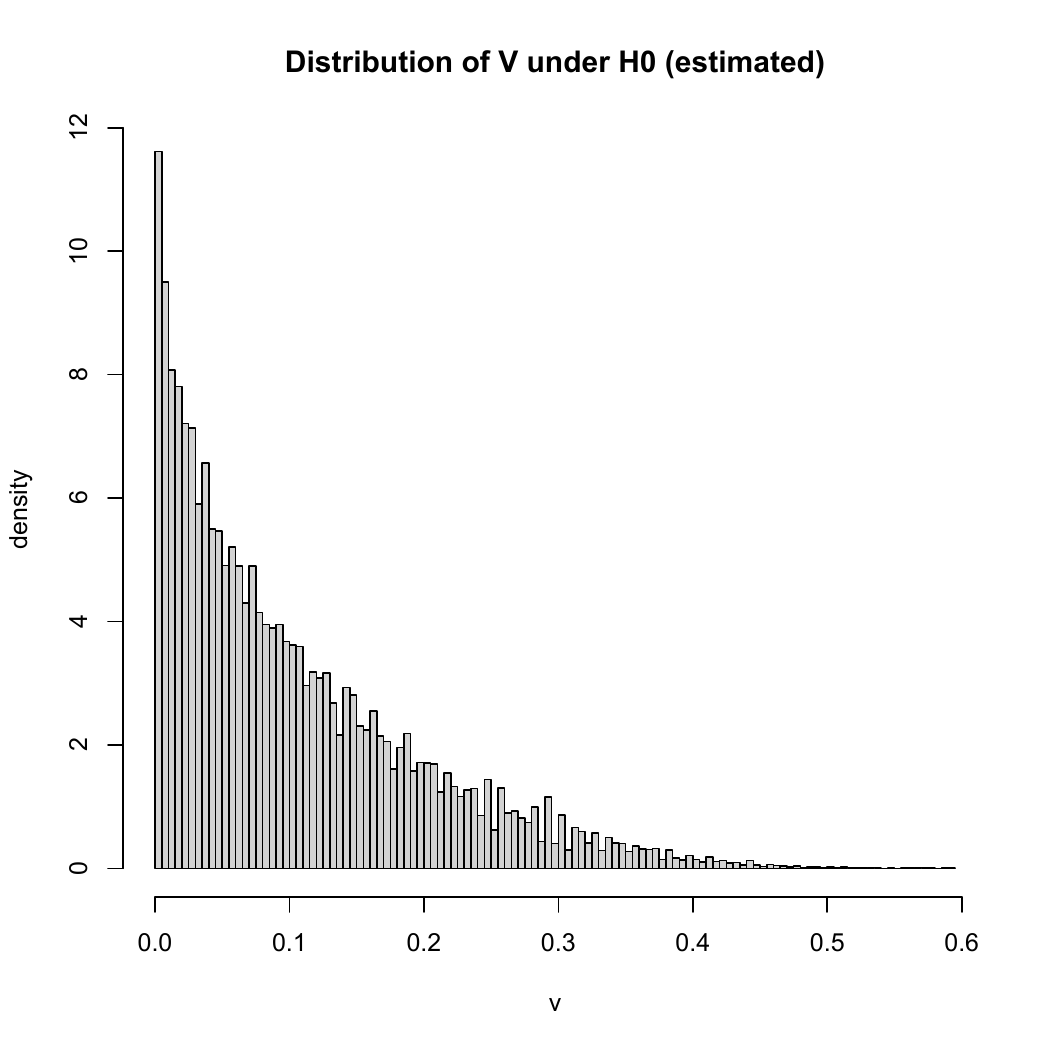}
    \caption{Empirical distribution of the test statistic $V$ defined by (\ref{eq:stattest}).  
Results obtained by simulating $10^5$ samples of size $n=57$.}
    \label{fig:loiVH0n}
\end{figure}

By simulating \lstinline|Nsim=10^5| i.i.d. samples of the standard normal distribution (or any other continuous distribution),
we obtain \lstinline|Nsim| realizations $v$ of $V$ and thus an empirical distribution representing the distribution of the statistic $V$ under the null hypothesis $H_0$ (no change-point)
which is displayed as an histogram in Figure~\ref{fig:loiVH0n}.

\blst
set.seed=123
Nsim=10^5 
X = matrix(rnorm(n * Nsim), nrow = Nsim, ncol = n) #Nsim white noise as rows 
clust=makeCluster(nclust)
clusterExport(clust,c("pvWMW","pvsWMW","v"))
loiVH0n = parApply(clust, X, 1, v) #Nsim realizations of V under H0
stopCluster(clust)
\end{lstlisting}

\n{\it Remark --}
The statistical test studied in this note is distribution-free, as is the WMW test;
that is, the distribution of the statistic $V$ does not depend on the continuous distribution under
$H_0$ of the random variables $X_t$ forming the time series.
To illustrate this fact, one can, after using the standard normal distribution,
estimate the distribution of $V$ by simulating, for example, \lstinline|Nsim=10^5| i.i.d. samples of the uniform distribution over the interval $(0,1)$, and then compare the two resulting empirical distributions.

\blst
set.seed=123
Nsim=10^5 
X = matrix(runif(n * Nsim), nrow = Nsim, ncol = n)  # Nsim white noise as rows 
clust=makeCluster(nclust)
clusterExport(clust,c("pvWMW","pvsWMW","v"))
loiVH0u = parApply(clust, X, 1, v)  # Nsim realizations of V under H0
stopCluster(clust)
\end{lstlisting}

\begin{figure}[t]
    \centering
    \includegraphics[scale=0.5]{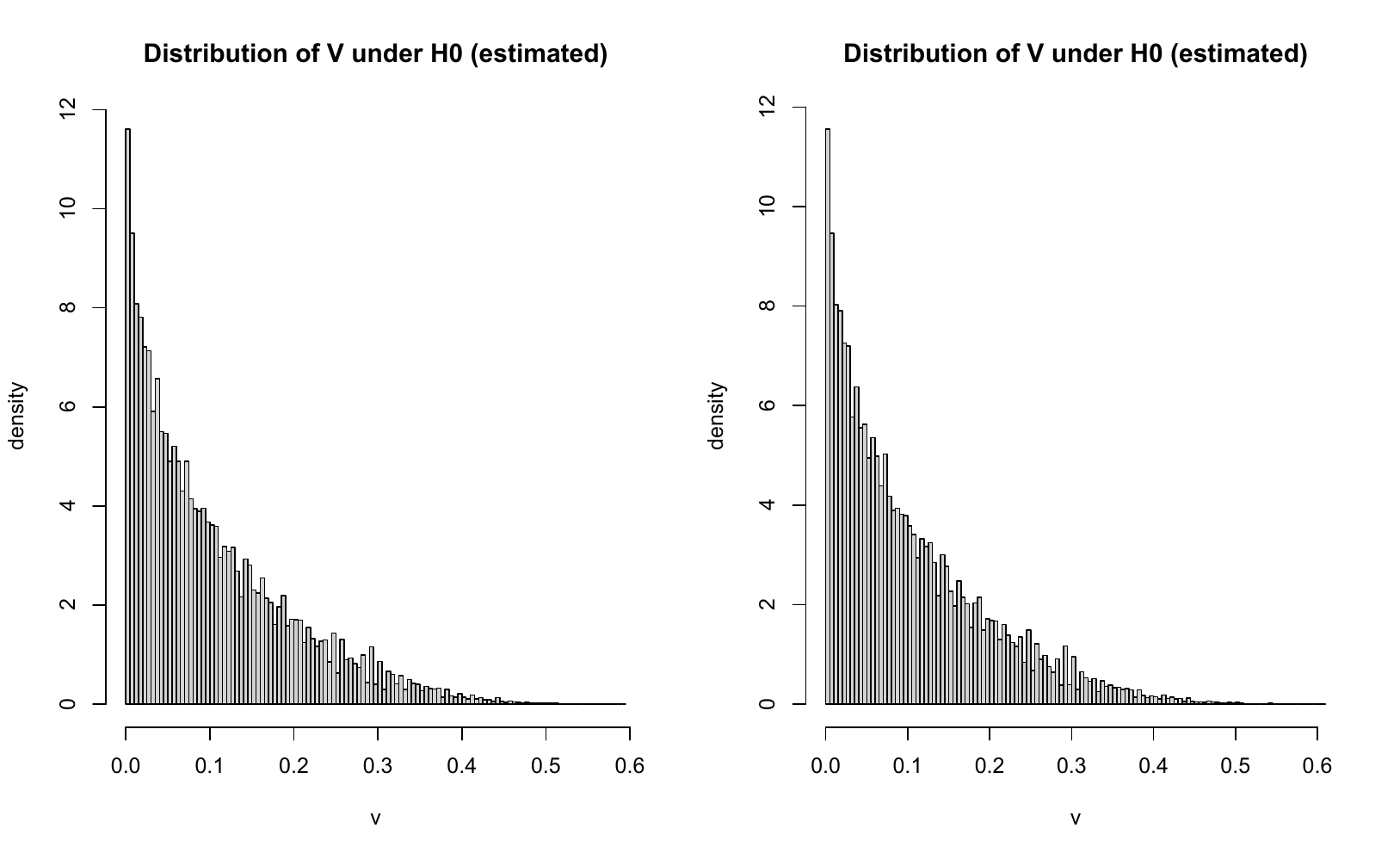}
    \caption{Empirical distributions of the test statistic $V$ defined by (\ref{eq:stattest}).
    Results obtained by simulating $10^5$ i.i.d. samples of size $n=57$ from the standard normal distribution (left panel) and the uniform distribution
    on $(0,1)$ (right panel).}
    \label{fig:loiVH0_n-u}
\end{figure}

\begin{figure}[t]
    \centering
    \includegraphics[scale=0.5]{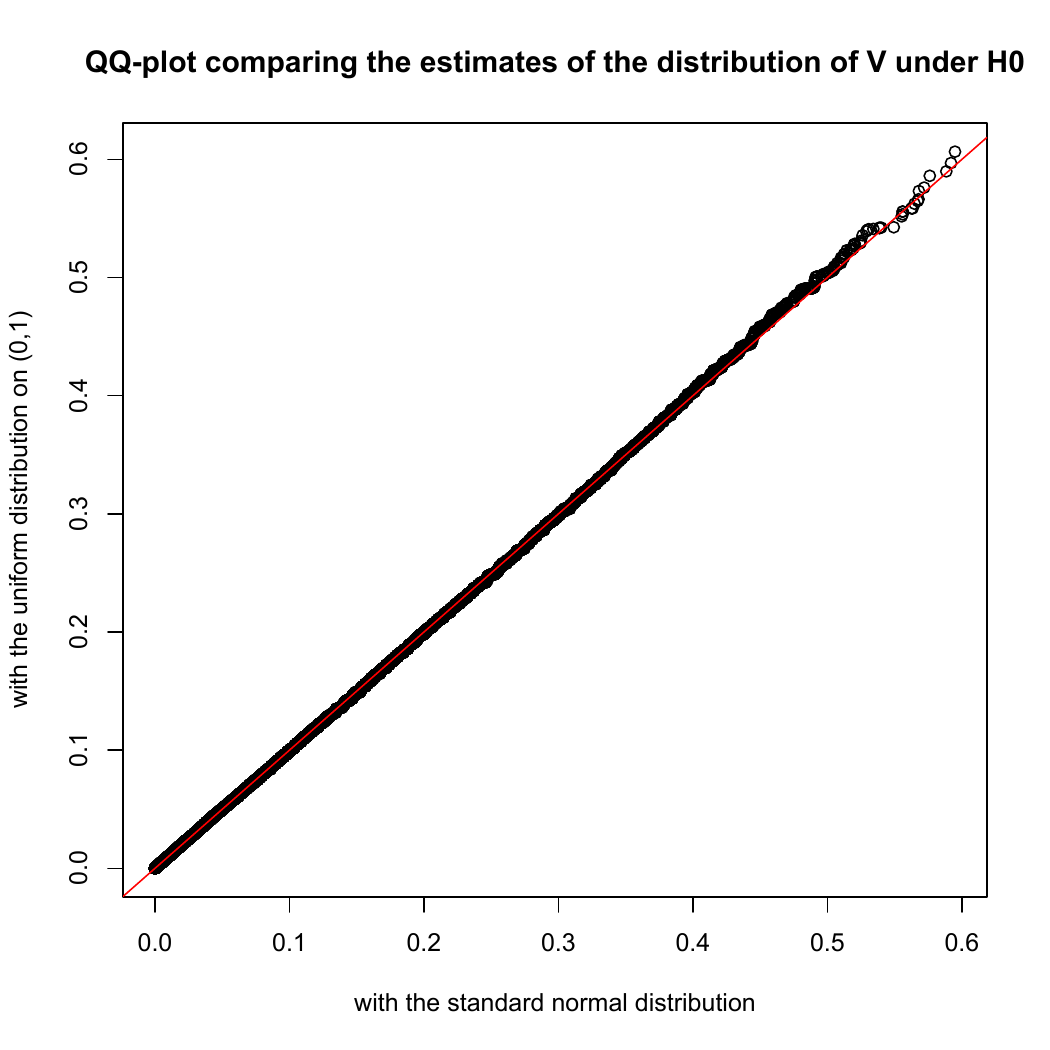}
    \caption{
Quantile-quantile plot for the empirical distributions of the test statistic $V$ defined by (\ref{eq:stattest}), obtained using the standard normal distribution and the uniform distribution on $(0,1)$.
}
    \label{fig:qqplot_loiVH0_n-u}
\end{figure}

In Figure~\ref{fig:loiVH0_n-u}, the histograms corresponding to these two empirical distributions are shown;
in Figure~\ref{fig:qqplot_loiVH0_n-u}, the corresponding quantile-quantile plot is displayed.
To obtain a more quantitative comparison, various statistical two-sample comparison tests can be applied:
using the Student’s t-test (\lstinline|t.test(loiVH0n, loiVH0u)|), we obtain a p-value of 0.3755522;
using the Wilcoxon-Mann-Whitney test (\lstinline|wilcox.test(loiVH0n, loiVH0u)|), we obtain a p-value of 0.519417;
using the Kolmogorov-Smirnov test (\lstinline|ks.test(loiVH0n, loiVH0u)|), we obtain a p-value of 0.2852201.

All of this is consistent with the non-parametric nature of the change-point detection test under study.

\section{Estimation of the $p$-value of the change-point detection test and estimation of the change-point}

Given a time series \lstinline|x|, the function \lstinline|detect_change_point| below returns  
an estimate of the $p$-value of the change-point detection test based on the empirical distribution of the test statistic  $V$ under $H_0$ obtained in the previous section,  
as well as an estimate of the possible change-point time given by the time corresponding to the minimum of the sequence of $p$-values  
from the successive WMW tests performed.

\blst
detect_change_point=function(x,loiVH0,b=6){
  pvs=pvsWMW(x)
  return(c(sum(loiVH0<min(pvs))/length(loiVH0),which.min(pvs)+(b-1)))
}
\end{lstlisting}

\v2
\n{\it Example --}
We simulate a time series \lstinline|x| of length $n=57$ from Gaussian distributions with variance 1,  
with a change-point at time $\tau=20$. The difference in means between the two blocks is equal to 1.  
For this time series \lstinline|x|, we determine the $p$-value of the test  
as well as an estimate of the change-point time.

The following instructions
\blst
tau = 20 #change-point
m1 = 1 #mean difference before/after the change
x = c(rnorm(tau), m1 + rnorm(n - tau))
detect_change_point(x, loiVH0n) #test for change-point (p-value,estimated change time)
\end{lstlisting}
returns in our example a $p$-value lower than $1\%$ and a change-point estimate close to $\tau = 20$, which seems to confirm the effectiveness of the test on the simulated sequence.

\blst
[1]  0.00824 18.00000
\end{lstlisting}

\begin{figure}[t]
    \centering
    \includegraphics[scale=0.5]{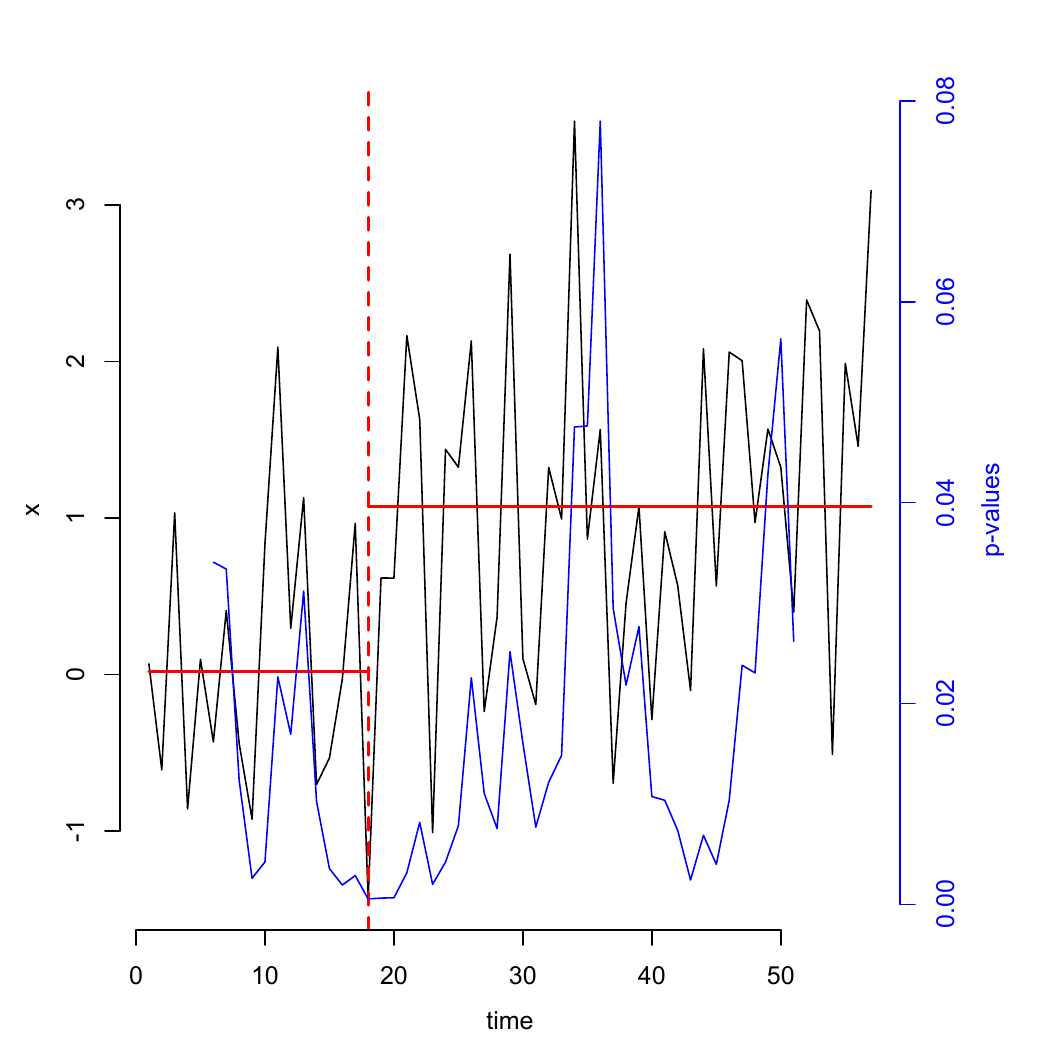}
    \caption{
Example of a time series with a change-point at time $\tau=20$, following Gaussian distributions: $\NR(0,1)$ before $\tau$ and $\NR(1,1)$ after $\tau$. Series of $p$-values from WMW tests (in blue). Estimated change-point $\hat{\tau}$ (dashed line) and empirical medians before and after $\hat{\tau}$ (in red).
}
    \label{fig:Ex1}
\end{figure}

We can plot on the same graph (Figure~\ref{fig:Ex1}) the time series \lstinline|x|,  
the sequence of $p$-values given by the successive WMW tests,  
the estimated change-point time $\hat{\tau}$,  
as well as the empirical medians of the values of the time series before and after $\hat{\tau}$.

\newpage
\section{On the power of the studied test}

After specifying hypothesis $H_1$, it is possible to estimate the power of the test using simulations. For that, \lstinline|Nsim2=10^4| sequences are simulated under $H_1$,  and the power is estimated  
by calculating the proportion of cases where a change-point was detected at the $5\%$ significance level.

Each configuration under $H_1$ considered below is characterized by a change-point $\tau$ (often chosen to be 20) and a one-parameter family of distributions governing the time series. The change is due to a shift in the parameter of the distribution family.

\subsection{Test power depending on the configurations under $H_1$}

In this subsection, we aim to illustrate how the power of the change-point detection test based on (\ref{eq:stattest}) depends on the configuration chosen under $H_1$.

To this end, we first consider configurations under $H_1$ based on the family of normal distributions with unit variance. Before the change-point $\tau=20$, the random variables $X_t$ in the time series follow the standard normal distribution; after $\tau$, they follow a normal distribution with mean $m_1$ and unit variance. The parameter $m_1$ is varied from 0 to 2 in steps of 0.1.

We also consider the family of uniform distributions over intervals of length four\footnote{so as to ensure a variance fairly close to one}, where the mean is similarly varied from 0 to 2 in steps of 0.1. More precisely, before the change-point $\tau=20$, we use the uniform distribution over the interval $(0,4)$;
after $\tau$, we use the uniform distribution over the interval $(0,4)+m_1$, with the same values of $m_1$ as above.

\begin{figure}[t]
    \centering
    \includegraphics[scale=0.5]{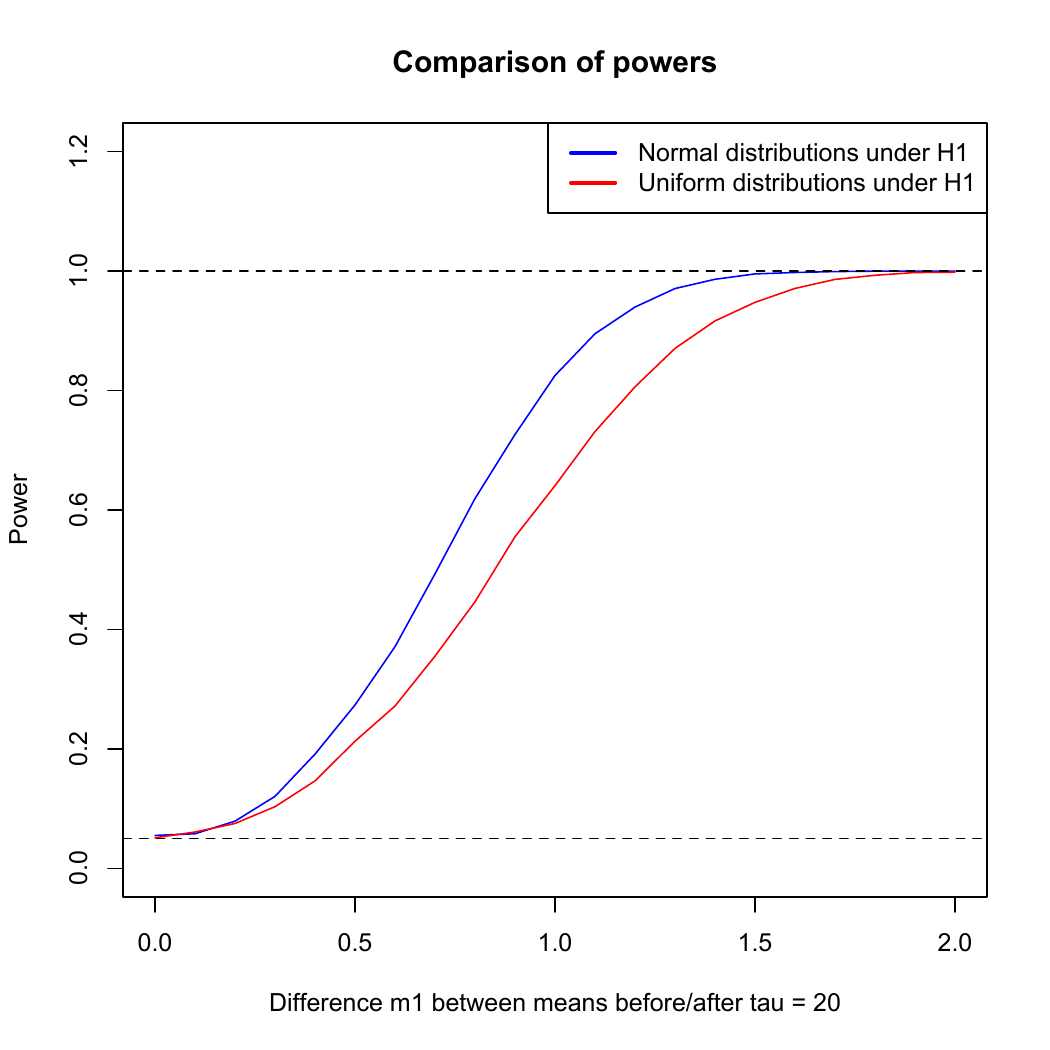}
    \caption{
Estimated test power under various configurations under $H_1$.
}
    \label{fig:puiss_config}
\end{figure}

In the following code, we estimate the power of the test for these different configurations under $H_1$. Results are shown in Figure~\ref{fig:puiss_config}. As expected, the power of the test increases with $m_1$ from $0.05$ when $m_1=0$ to $1$ for larger values of $m_1$. Similarly, using normal distributions with smaller variance, or uniform distributions over narrower intervals, would result in higher test power.

\blst
clust=makeCluster(nclust)
clusterExport(clust, c("n","pvWMW","pvsWMW","loiVH0n","detect_change_point"))
tau=20
Nm1=21
puissancesH1n = numeric(Nm1)
puissancesH1u = numeric(Nm1)
Nsim2=10^4
set.seed=123
for (i in 1:Nm1){
    m1 = (i-1)/10
    XH1n=cbind(matrix(rnorm(tau*Nsim2),nrow=Nsim2),matrix(m1+rnorm((n-tau)*Nsim2),nrow=Nsim2))
    XH1u=cbind(matrix(runif(tau*Nsim2,0,4),nrow=Nsim2),matrix(m1+runif((n-tau)*Nsim2,0,4),nrow=Nsim2))
    SH1n=parApply(clust,XH1n,1,detect_change_point,loiVH0n)
    SH1u=parApply(clust,XH1u,1,detect_change_point,loiVH0n)
    puissancesH1n[i] = sum(SH1n[1,]<.05)/Nsim2
    puissancesH1u[i] = sum(SH1u[1,]<.05)/Nsim2
}
stopCluster(clust)
\end{lstlisting}

\subsection{Comparison with alternative change-point methods based on the WMW test}

Pettitt’s change-point detection test \cite{Pettitt} shares similarities with the test discussed in this note. It consists of applying,  
at each time $\tau$ of the considered time series \lstinline|x|,  
the WMW test to compare the samples \lstinline|x[1:tau]| and \lstinline|x[(tau+1):n]|.  
The Pettitt test statistic is then defined as the maximum of the absolute values  
of the centered WMW test statistics. An alternative approach discussed in \cite{Zhou} consists in standardizing the  WMW test statistics before taking the maximum of the absolute values.  
In other words, denoting by $U_\tau$ the WMW test statistic corresponding to the  
comparison of \lstinline|x[1:tau]| and \lstinline|x[(tau+1):n]|, the Pettitt test statistic is given by
\begin{equation}
\label{eq:Pettitt}
W=
\max_{b \leq \tau \leq n-b} \left\{ \left| U_\tau - E(U_\tau)\right| \right\}\,
\end{equation}
whereas its standardized alternative is defined as
\begin{equation}
\label{eq:Pettitts}
W\!s=
\max_{b \leq \tau \leq n-b} \left\{ \left| \frac{U_\tau - E(U_\tau)}{\sqrt{\mathrm{var}(U_\tau)}} \right| \right\}\,.
\end{equation}

\blst
f=function(k,x,stand=TRUE){
  #compute the test statistics based on WMW
  #stand = TRUE : standardized version
  #stand = FALSE :non-standardized version of Pettitt 
  n1 = k
  n2 = length(x)-k
  E = n1*n2 / 2  #mean of the test statistic under H0
  S = sqrt(n1*n2*(n1+n2+1)/12) #variance of the test statistic under H0
  if (stand){ 
    return( abs( (wilcox.test(x[1:k],x[(k+1):(length(x))])$statistic - E)/S ) )
  }else{  #non standardized version, Pettitt version
    return( abs( wilcox.test(x[1:k],x[(k+1):(length(x))])$statistic -E ) )
  }  
  }

f2=function(x,b=6,stand=TRUE){  
  stats=sapply(b:(length(x)-b),f,x=x,stand=stand)
  return(c(max(stats),which.max(stats)+b-1))
}

estim_rupture=function(x,loiH0,stand=TRUE){
  tst=f2(x,stand=stand)
  return(c(sum(loiH0>tst[1])/length(loiH0),tst[2]))
}

#Compute empirical distribution of the test statistic under H0, standardized version
Nsim=10^5
X=matrix(rnorm(n*Nsim),nrow=Nsim,ncol=n)
clust <- makeCluster(nclust)
clusterExport(clust,c("f","f2"))
S=parApply(clust,X,1,f2,stand=TRUE)
stopCluster(clust)
loiPsH0=S[1,]  

#Compute empirical distribution of the test statistic under H0, non-standardized version
clust <- makeCluster(nclust)
clusterExport(clust,c("f","f2"))
S=parApply(clust,X,1,f2,stand=FALSE)
stopCluster(clust)
loiPH0=S[1,]

#Estimate power of the tests using simulations
clust=makeCluster(nclust)
clusterExport(clust, c("n","pvWMW","pvsWMW","loiVH0n","detect_change_point","f","f2","estim_rupture","loiPsH0","loiPH0"))
tau=20
Nm1=21
puissancesH1n = numeric(Nm1)  #using p-value
puissancesPsH1n = numeric(Nm1) #using standardized Pettitt
puissancesPH1n = numeric(Nm1) #using original Pettitt

Nsim2=10^4
for (i in 1:Nm1){
  m1 = (i-1)/10
  XH1n=cbind(matrix(rnorm(tau*Nsim2),nrow=Nsim2),matrix(m1+rnorm((n-tau)*Nsim2),nrow=Nsim2))
  SH1n=parApply(clust,XH1n,1,detect_change_point,loiVH0n)  #using p-value
  SPsH1n=parApply(clust,XH1n,1,estim_rupture,loiPsH0,stand=TRUE)  #using standardized Pettitt
  SPH1n=parApply(clust,XH1n,1,estim_rupture,loiPH0,stand=FALSE) #using original Pettitt
  puissancesH1n[i] = sum(SH1n[1,]<.05)/Nsim2
  puissancesPsH1n[i] = sum(SPsH1n[1,]<.05)/Nsim2
  puissancesPH1n[i] = sum(SPH1n[1,]<.05)/Nsim2
}
stopCluster(clust)
\end{lstlisting}

\begin{figure}[t]
    \centering
    \includegraphics[scale=0.5]{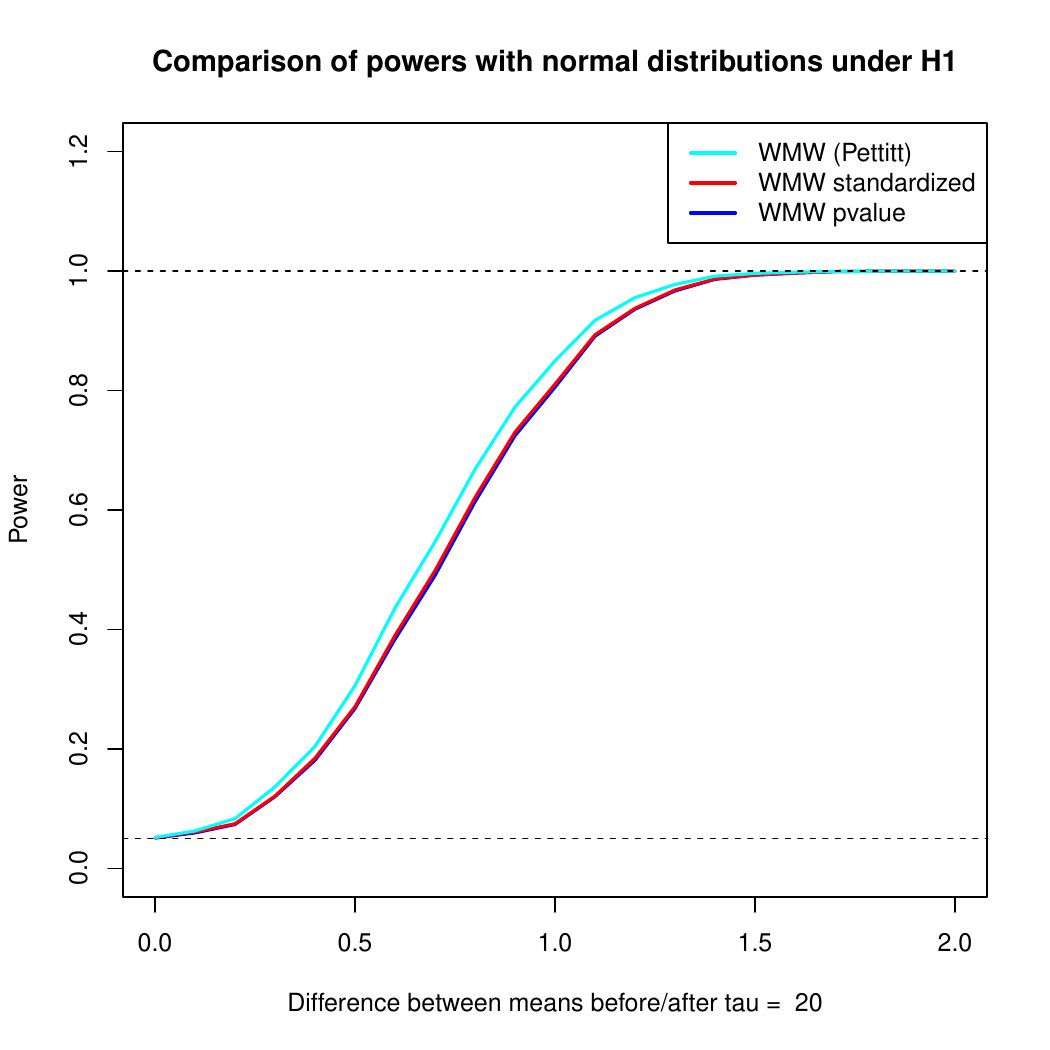}
    \caption{Power of three change-point detection procedures based on the WMW test statistic as a function of the difference in mean between the two samples. Cyan [resp. red; blue] curve corresponds to the test statitics defined by (\ref{eq:Pettitt}) [resp. (\ref{eq:Pettitts}) ; (\ref{eq:stattest})]
 Results obtained by simulating $10^4$ samples of size $n=57$ and change-point time $\tau =20$.}
    \label{fig:compPettitt}
\end{figure}

Figure~\ref{fig:compPettitt} shows that, as expected, the tests based on the normalized statistics  (\ref{eq:Pettitts}) and (\ref{eq:stattest}) have a similar power for the experimental setting considers in this study. A less expected result is that the test based on the non-normalized test statistic (\ref{eq:Pettitt}) provides a more powerful test compared to the two normalized versions. 

\begin{figure}[t]
    \centering
    \includegraphics[scale=0.5]{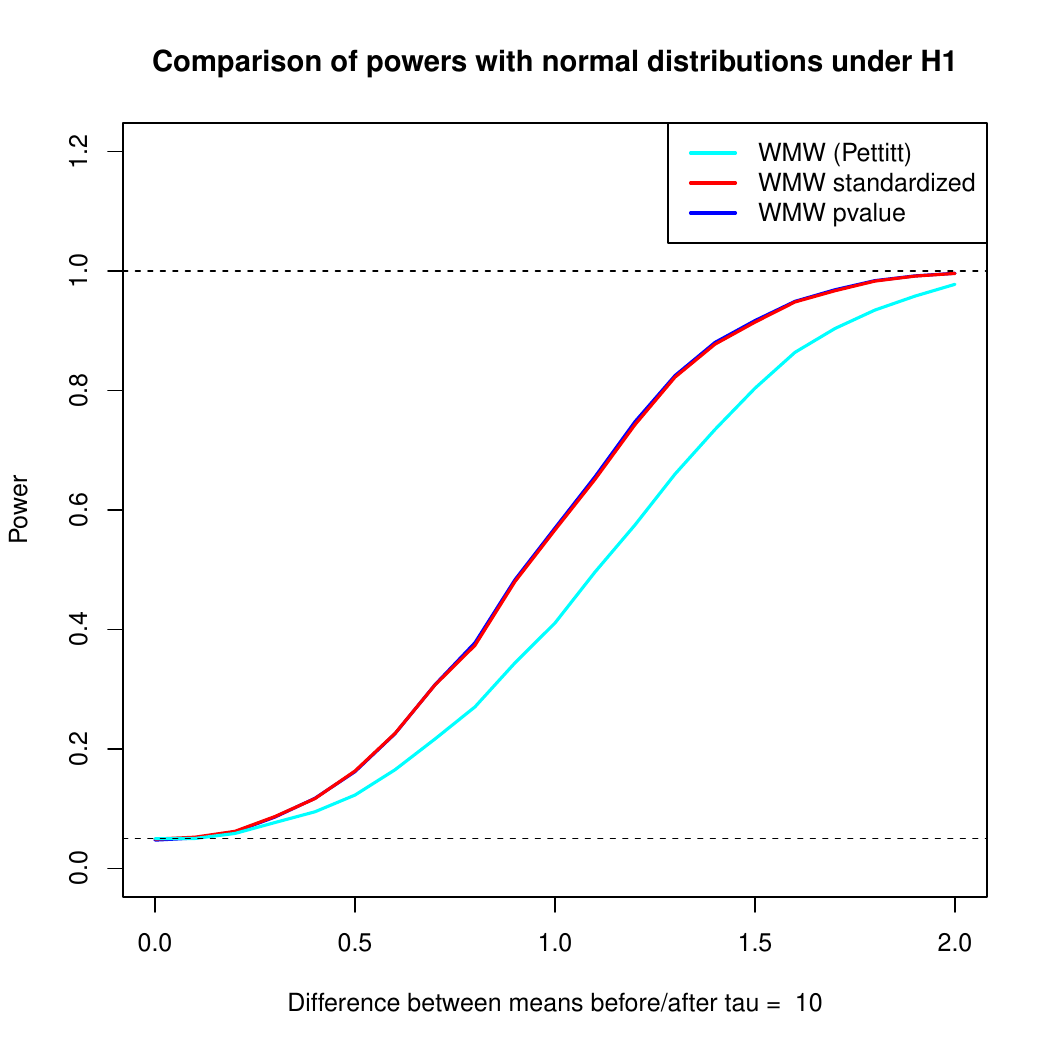}
    \caption{Same as Figure~\ref{fig:compPettitt}  with $\tau =10$.}
    \label{fig:compPettitt10}
\end{figure}

To further investigate this result, the experiment was run again with $\tau=10$ instead of $\tau=20$, see Figure~\ref{fig:compPettitt10}. Again, the two tests based on the normalized statistics have similar powers, but become clearly more powerful than the test based on the non-normalized test statistic. Further experiments with different values of $\tau$ were performed, it was found that the non-normalized test statistic (\ref{eq:Pettitt}) is slightly more powerful when the change-point $\tau$ is close to the middle of the sequence (i.e. $\tau \approx \frac n 2$), but the normalized alternatives are more powerful otherwise. A possible explanation is that the normalisation will reduce the sensibility of the test statistic when $\tau$ is close to $n/2$ since 
$\mathrm{var}(U_\tau)=\tau(n-\tau)(n+1)/12$ is maximum for $\tau=\frac n 2$.

\subsection{Comparison with alternative change-point methods based on the Student’s t-test}

The framework described in the previous sections can be used in conjunction with any other two samples comparison test than the WMW test.
In this section,
We consider two alternatives based on the repeated use of Student’s t-tests (assuming equal variances):
one using the statistic given by the maximum of the absolute values of the statistics of the t-tests performed,
the other using the minimum of the $p$-values of these same tests.

Once again, the possibility of a change-point in the first five and last five time points is excluded,  
and both cases of normal and exponential distributions under hypothesis $H_1$ are considered.

\subsubsection{Case of normal distributions under $H_1$}

\begin{figure}[t]
    \centering
\includegraphics[scale=0.4]{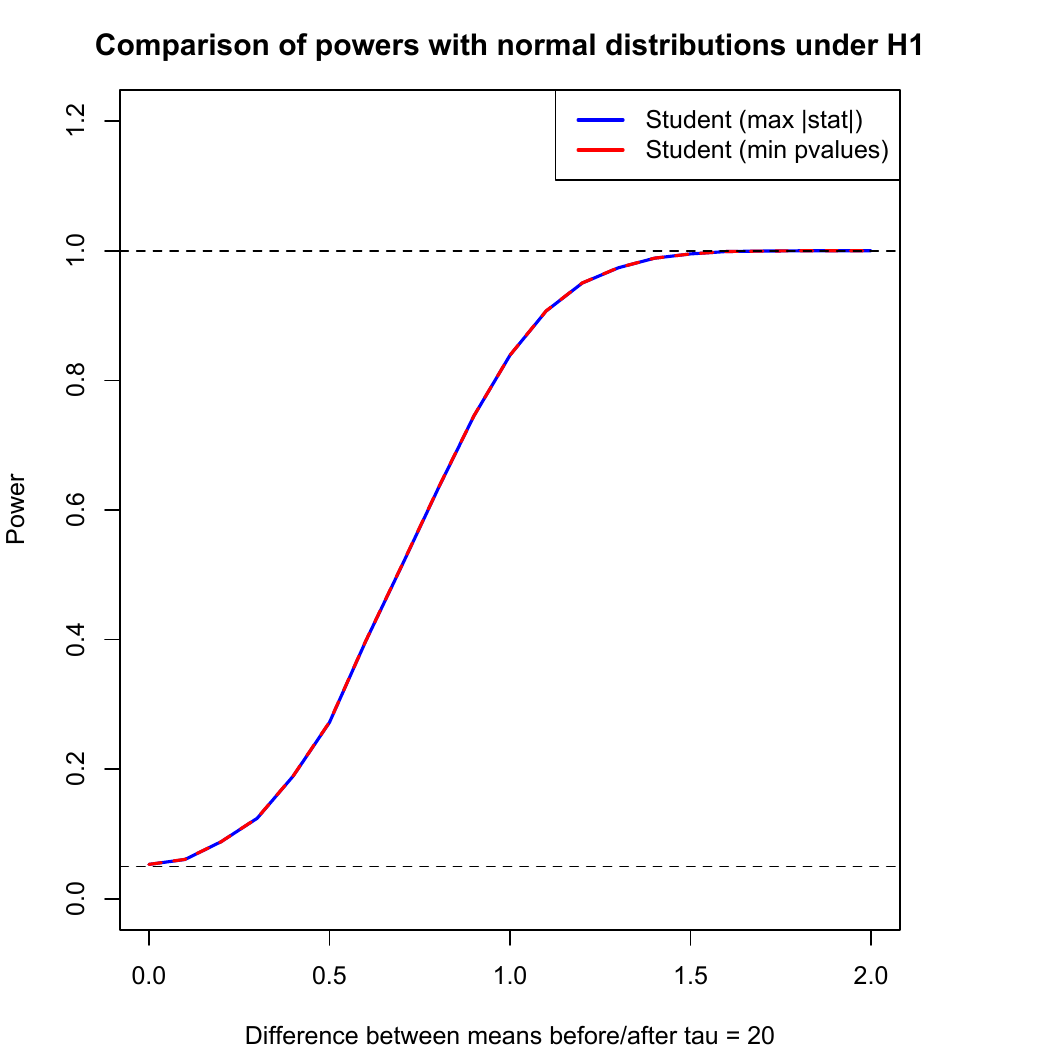}
\includegraphics[scale=0.4]{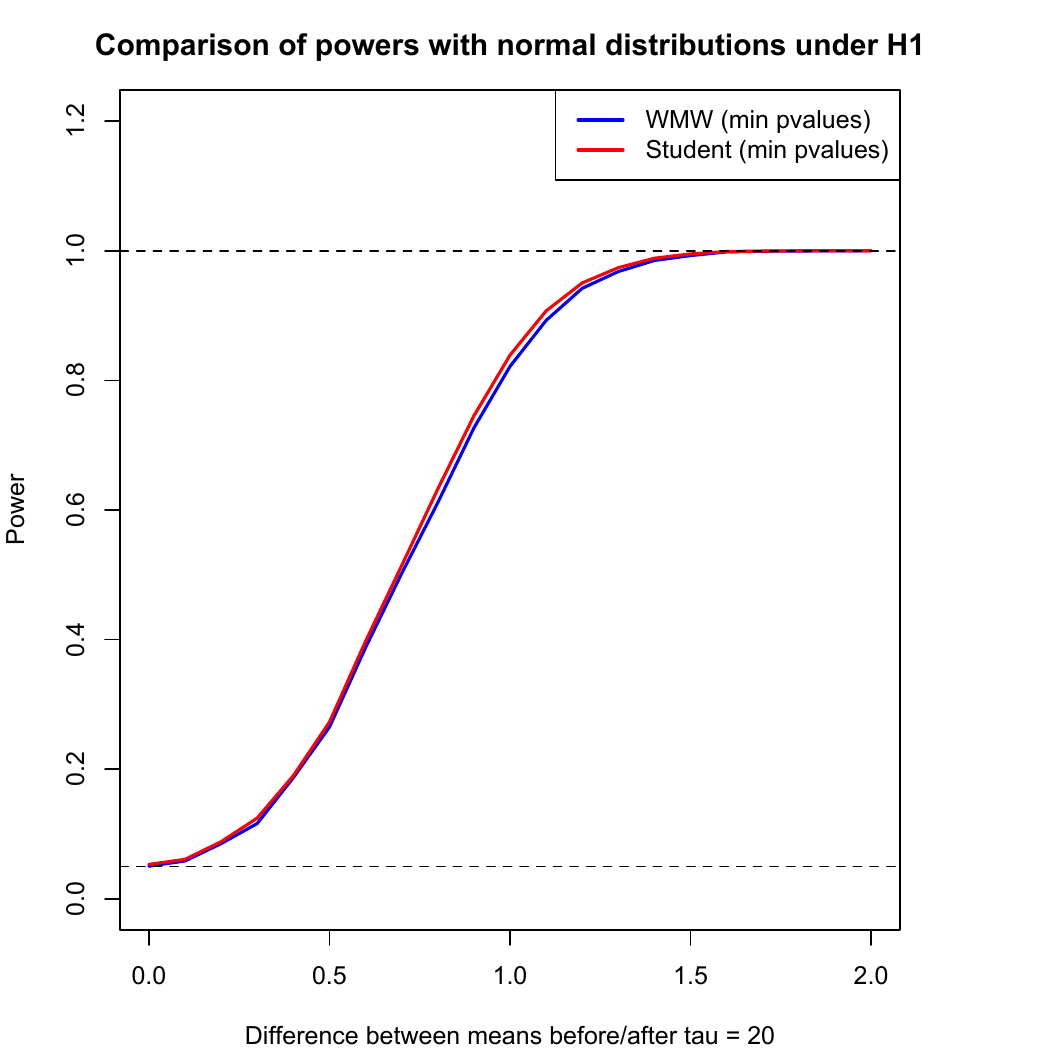}
  \caption{
Power of three change-point detection tests as a function of the mean difference across the change-point.
Left panel: tests based on repeated Student's $t$-tests --
Test 1 (in blue) using the maximum absolute value of the $t$-statistics,
and Test 2 (in red) using the minimum of the $p$-values.
Right panel:
Test 2 (in red), and Test 3 (in blue) using (\ref{eq:stattest}).
Results obtained by simulating $10^4$ time series of length $n=57$, following normal distributions, with a change-point at $\tau=20$.
}
    \label{S_S_WMWn}
\end{figure}

First, we note that, with normal distributions under $H_1$, the two tests based on the Student’s t-test exhibit similar performance
(Figure~\ref{S_S_WMWn}, left).

We then compare  the
two tests based on the minimum of the $p$-values,
using WMW tests for one and Student’s t-tests for the other.  
In this case of normal distributions under $H_1$,  
the latter is slightly more powerful than the former
(Figure~\ref{S_S_WMWn}, right).

\subsubsection{Case of exponential distributions under $H_1$}

\begin{figure}[t]
    \centering
\includegraphics[scale=0.4]{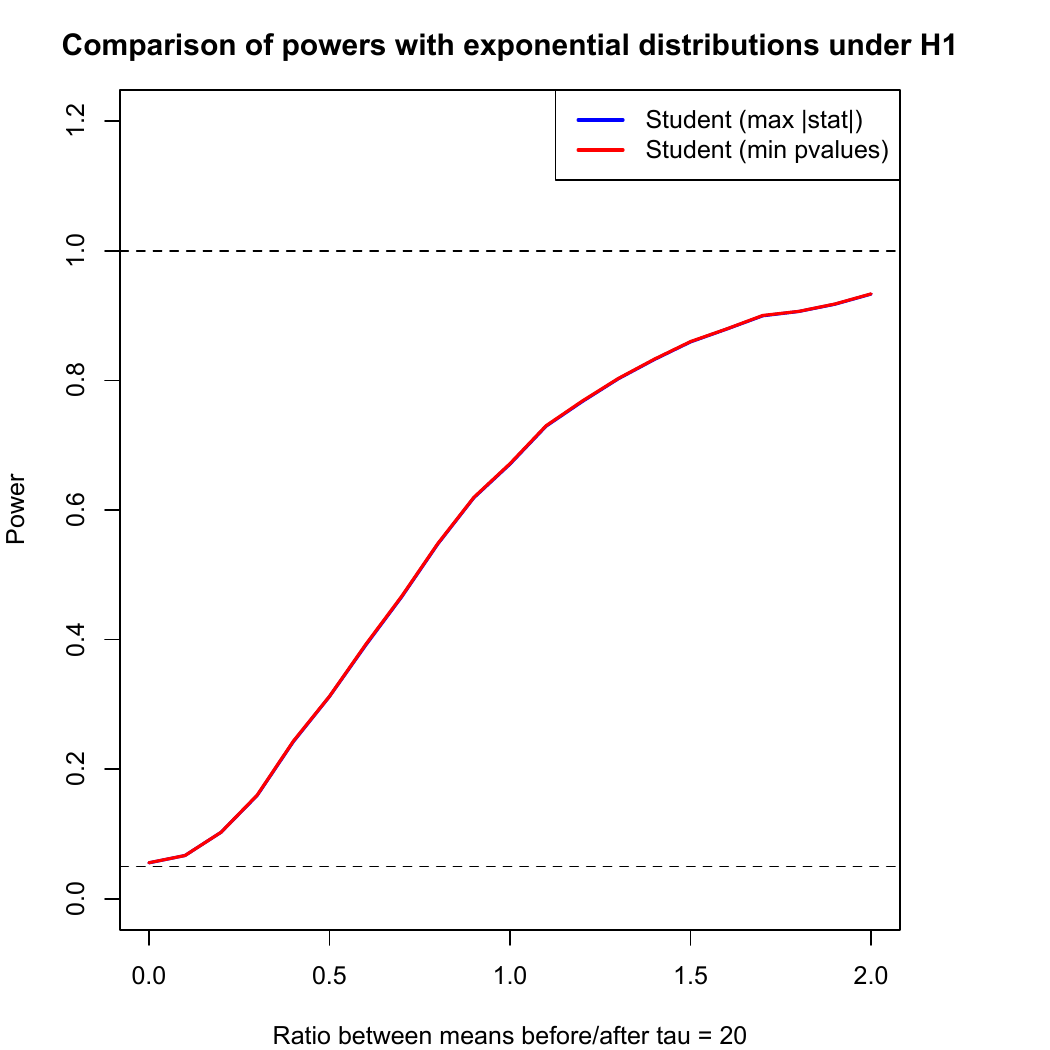}
\includegraphics[scale=0.4]{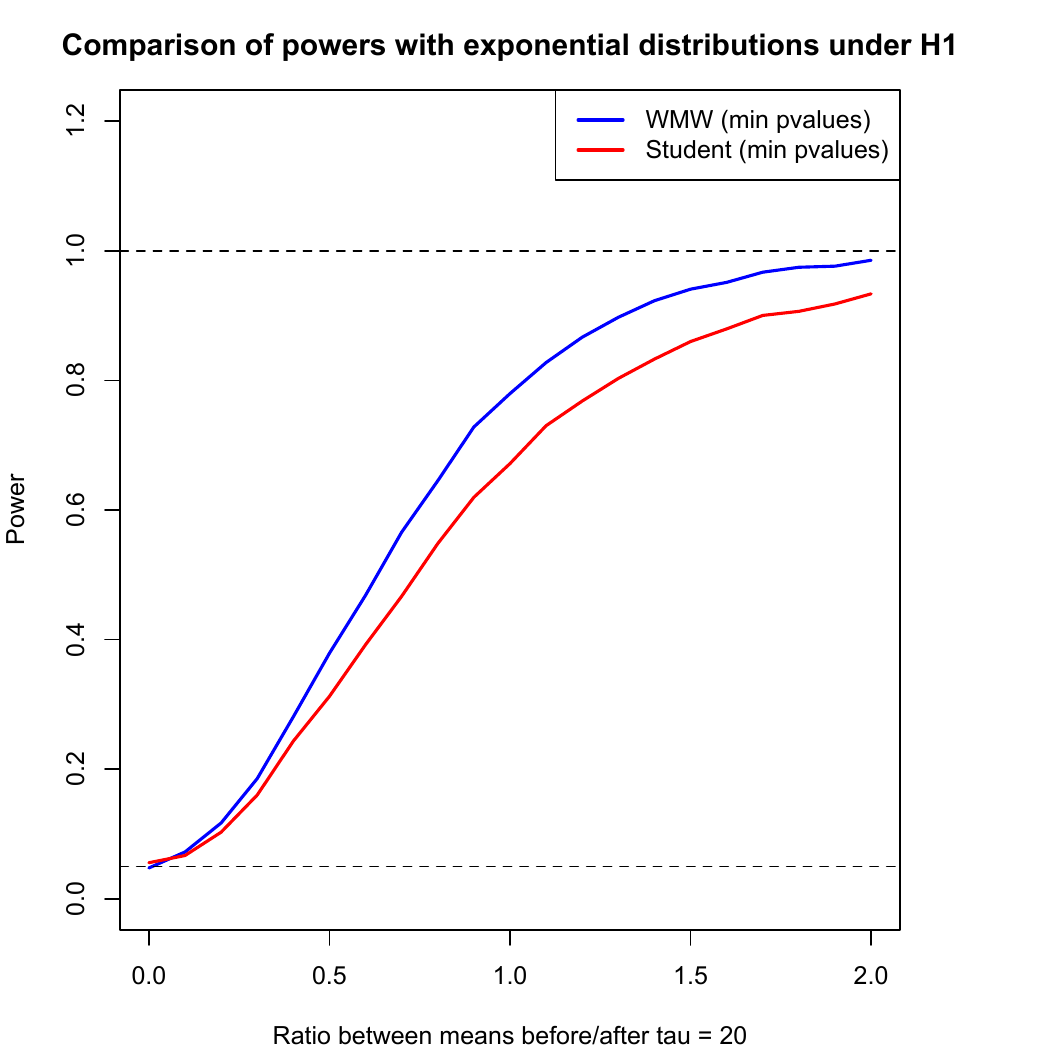}
  \caption{
Same as Figure~\ref{S_S_WMWn} with exponential distributions under $H_1$.
}
    \label{S_S_WMWe}
\end{figure}

In this case as well, the two tests based on the Student’s t-test are very similar
(Figure~\ref{S_S_WMWe}, left).
However, the test studied in this note proves to be more powerful than those using Student’s t-test in this situation
(Figure~\ref{S_S_WMWe}, right).  

\newpage
\subsection{Comparison with a likelihood ratio test}

When, under hypothesis $H_1$,  
we have normal distributions and the change-point  
occurs in the mean ($\mu_1 \neq \mu_2$) with constant variance ($\sigma^2$) at time $\tau$,  
the log-likelihood function under $H_1$ is given by:
$$
\ell_{H_1,\tau} (\mu_1,\mu_2,\sigma;x)
=
-n\ln(\sigma\sqrt{2\pi}) - \frac{1}{2\sigma^2}\left(\sum_{i=1}^\tau (x_i-\mu_1)^2 + \sum_{i=\tau+1}^n (x_i-\mu_2)^2\right).
$$
It is maximized when the triplet $(\mu_1, \mu_2, \sigma^2)$ equals
$$
\left(\overline{x_{\tau,1}} = \frac{1}{\tau} \sum_{i=1}^\tau x_i, \quad
\overline{x_{\tau,2}} = \frac{1}{n-\tau} \sum_{i=\tau+1}^n x_i, \quad
s_{\tau}^2 = \frac{1}{n} \left(\sum_{i=1}^\tau (x_i - \overline{x_{\tau,1}})^2 + \sum_{i=\tau+1}^n (x_i - \overline{x_{\tau,2}})^2 \right) \right).
$$
Hence,
$$
\max_{\mu_1,\mu_2,\sigma} \ell_{H_1,\tau} (\mu_1,\mu_2,\sigma;x)
= -\frac{n}{2} \ln(2\pi s_{\tau}^2) - \frac{n}{2}.
$$
The analogous calculation under hypothesis $H_0$ gives:
$$
\max_{\mu,\sigma} \ell_{H_0} (\mu,\sigma;x)
= -\frac{n}{2} \ln(2\pi s^2) - \frac{n}{2},
$$
with $s^2 = \overline{(x - \overline{x})^2}$.

We thus deduce the log of the maximized likelihood ratio:
$$
\lambda(x) = \max_{1 \leq \tau \leq n-1} \left\{ \frac{n}{2} \left( \ln(s^2) - \ln(s_{\tau}^2) \right) \right\} \quad (\geq 0).
$$

Again, to reasonably compare  
the test studied in this note  
with  
the test associated to this likelihood ratio,  
one must assume that the change-point cannot occur at the beginning or end of the time series, and therefore consider the test statistic:
\begin{equation}
\label{eq:lrtest}
\lambda(x) = \max_{b \leq \tau \leq n - b} \left\{ \frac{n}{2} \left( \ln(s^2) - \ln(s_{\tau}^2) \right) \right\}
\quad
\text{with } b=6.
\end{equation}

\begin{figure}[t]
    \centering
\includegraphics[scale=0.4]{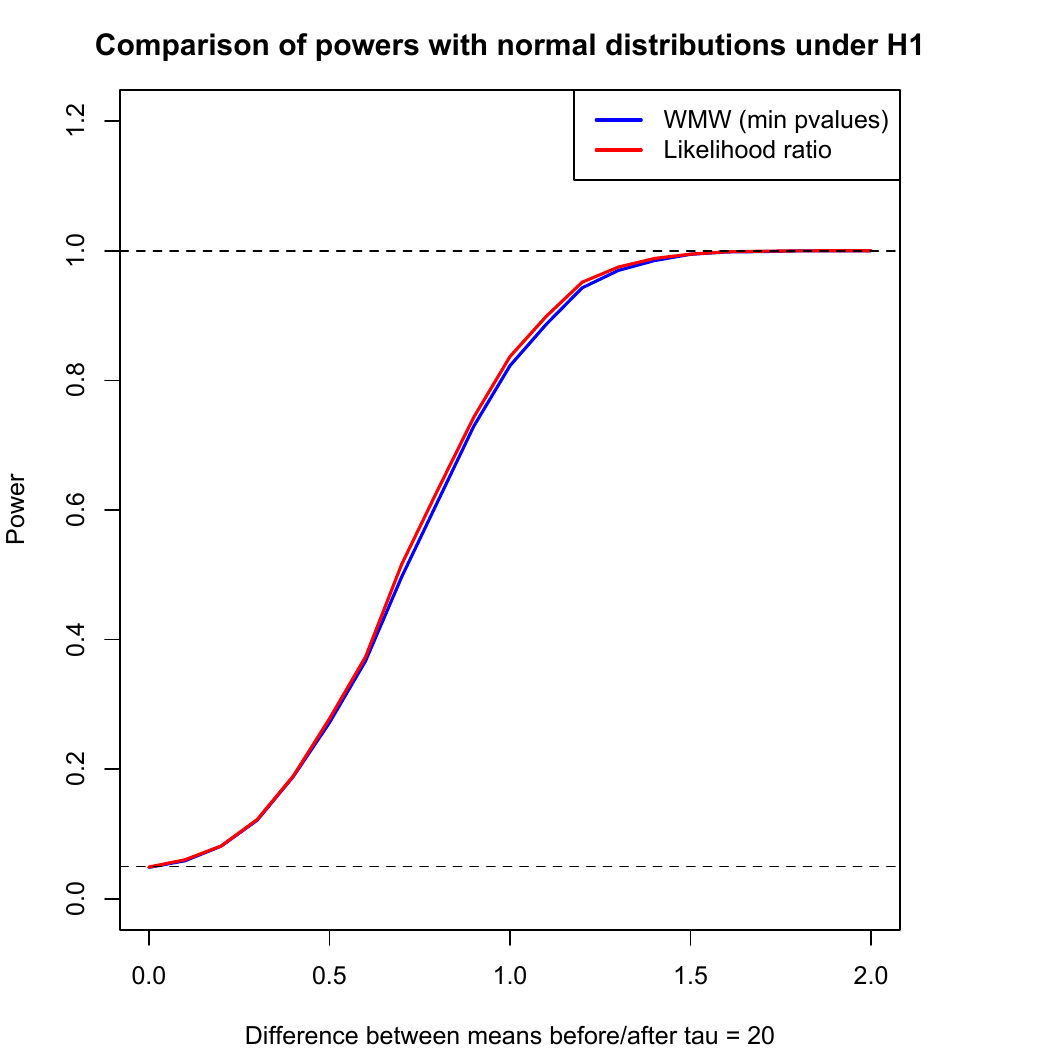}
\includegraphics[scale=0.4]{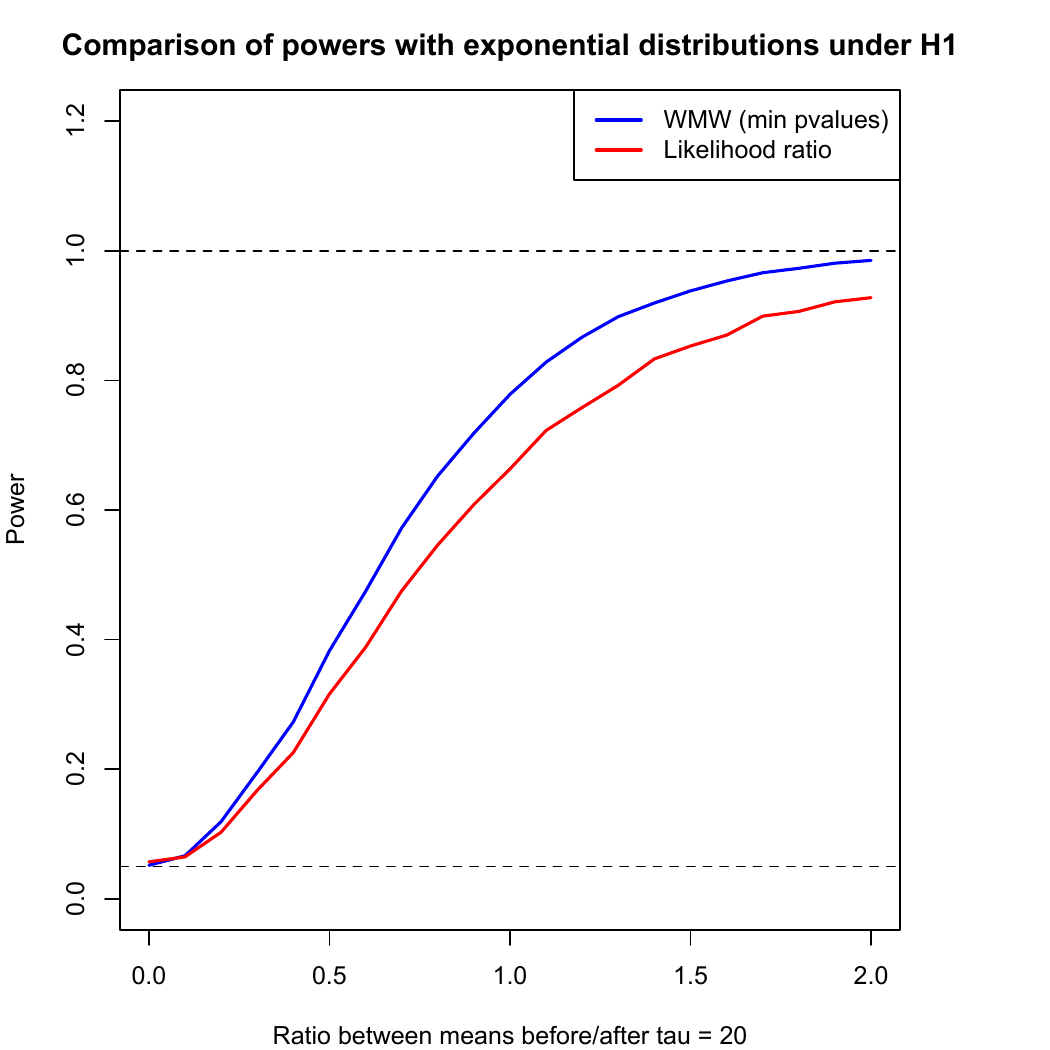}
  \caption{
Power of two change-point detection tests as a function of the mean difference/ratio across the change-point.
Left panel: Test defined by (\ref{eq:stattest}) (in blue) versus the likelihood ratio test defined by (\ref{eq:lrtest}) (in red).
Results obtained by simulating $10^4$ time series of length $n=57$, following normal distributions, with a change-point at $\tau=20$.
Right panel: Same as left panel, but with exponential distributions.
}
    \label{WMWvsRV}
\end{figure}

Figure~\ref{WMWvsRV} show that the use of this test statistic  
yields a slightly more powerful test when the distributions under $H_1$ are indeed normal, but  
less powerful for exponential distributions under $H_1$.

\section{Determination of a confidence interval for the estimation of the change-point}

\subsection{Two functions for determining confidence intervals}

The two functions \lstinline|ICboot1| and \lstinline|ICboot2| below  
each allow the determination of a confidence interval for the estimation of a possible change-point using a resampling technique (see \cite{Efron}, for example, for an introduction to these techniques).

Each of these two functions takes as input a time series \lstinline|x|  
and  
starts by determining the point estimate  
$\hat{\tau}$ of the change-point corresponding to the minimum of the $p$-values of the successive WMW tests applied to \lstinline|x|.

In the function \lstinline|ICboot1|,  
we then simulate \lstinline|Nboot=10^3| time series whose first $\hat{\tau}$ values (resp. last $n-\hat{\tau}$ values) are randomly drawn with replacement  
from the first $\hat{\tau}$ values (resp. last $n-\hat{\tau}$ values) of the original time series \lstinline|x|.  
For each of these $10^3$ new time series, a change-point is estimated as was done with $\hat{\tau}$ for the series \lstinline|x|.  
The 0.025 and 0.975 quantiles of the series of these $10^3$ estimated change-points provide the endpoints of the confidence interval given by \lstinline|ICboot1|.

The same procedure is applied in the function \lstinline|ICboot2|,  
except that, to simulate each of the \lstinline|Nboot=10^3| time series by resampling,  
instead of $\hat{\tau}$, a time point $\hat{\tau}'$ is randomly chosen from the set $\{\hat{\tau}-1, \hat{\tau}, \hat{\tau}+1\}$.

We start by defining a function \lstinline|pvWMWmat|,  
similar to the function \lstinline|pvWMW|  
but which speeds up computations thanks to a matrix-based treatment of the WMW tests to be performed.

\blst
pvWMWmat = function(k, x){  
  return(row_wilcoxon_twosample(x[,1:k], x[,(k+1):ncol(x)])$pvalue)
}

ICboot1 = function(x, Nboot = 10^3, b = 6){
  htau = which.min(pvsWMW(x)) + (b - 1)  # point estimate of the change-point
  #simulation des echantillons re-echantillonnes:
  n = length(x)
  Xboot = cbind(
    matrix(sample(x[1:htau], size = htau*Nboot, replace = TRUE), nrow = Nboot),
    matrix(sample(x[(1+htau):n],size = (n - htau)*Nboot, replace = TRUE),nrow = Nboot)
  )
  # sequence of WMW p-values for each bootstrapped sample:
  pvs=sapply(b:(n-b),pvWMWmat,x=Xboot)
  # change-point estimate for each bootstrapped sample:
  Sboot=apply(pvs, 1, which.min) + (b - 1) 
  return(quantile(Sboot, c(0.025, 0.975)))
}

ICboot2 = function(x, Nboot = 10^3, b = 6) {
  htau = which.min(pvsWMW(x)) + (b - 1)  # point estimate of the change-point
  htau_extended = seq(htau - 1, htau + 1) 
  # simulation of bootstrapped samples:
  htauprimes = sample(htau_extended, size = Nboot, replace = TRUE)
  n = length(x)
  Xboot = matrix(1:(Nboot * n), nrow = Nboot)
  for (i in 1:Nboot) {
    y = c(
      sample(x[1:htauprimes[i]], size = htauprimes[i], replace = TRUE),
      sample(x[(1+htauprimes[i]):n], size = (n - htauprimes[i]), replace = TRUE)
    )
    Xboot[i, ] = y
  }
  # sequence of WMW p-values for each bootstrapped sample:
  pvs = sapply(b:(n - b), pvWMWmat, x = Xboot)
  # change-point estimate for each bootstrapped sample:
  Sboot = apply(pvs, 1, which.min) + (b - 1)
  return(quantile(Sboot, c(0.025, 0.975)))
}
\end{lstlisting}

\v5
\n{\it Illustration --}  
We begin by simulating a time series \lstinline|x|, a realization of independent random variables  
following normal distributions with variance 1,  
with a change-point at time $\tau=20$ caused by a mean difference  
equal to \lstinline|m1=1|.  

We then test whether a change-point is detected and obtain a point estimate  
of the change-point using the function \lstinline|detect_change_point|.

The functions \lstinline|ICboot1| and \lstinline|ICboot2| then provide two confidence intervals for the estimation  
of the change-point, from which one can, in particular, calculate the length.

\blst
m1=1
tau=20
x = c(rnorm(tau),m1+rnorm(n-tau))
 
detect_change_point(x,loiVH0n)
 
IC1=ICboot1(x)
IC1
IC1[2]-IC1[1]

IC2=ICboot2(x)
IC2
IC2[2]-IC2[1]
\end{lstlisting}

\begin{figure}[t]
    \centering
    \includegraphics[scale=0.5]{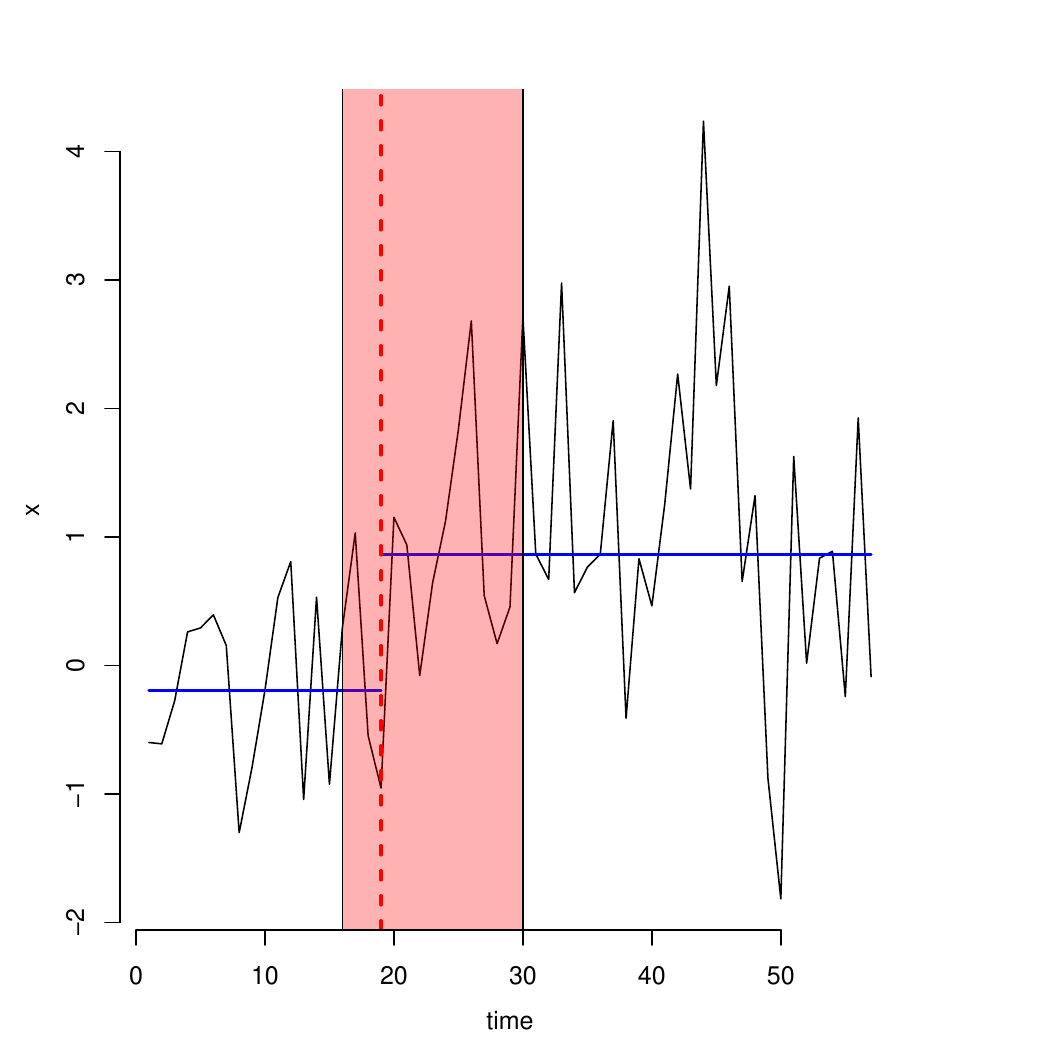}
    \caption{
Example of a time series with a change-point at time $\tau=20$, following Gaussian distributions: $\NR(0,1)$ before $\tau$ and $\NR(1,1)$ after $\tau$.  Estimated change-point $\hat{\tau}$ (dashed line) with associated confidence interval (red area) and empirical medians before and after $\hat{\tau}$ (in blue).
}. 
    \label{fig:Ex2}
\end{figure}

We can plot on the same graph (Figure~\ref{fig:Ex2}) a simulated time series with  
the estimated change-point time $\hat{\tau}$ and the confidence interval obtained using the function \lstinline|ICboot1|.

\subsection{Validation}

The aim of this section is to estimate by simulation the confidence level  
and the average length of the intervals obtained  
using the functions \lstinline|ICboot1| and \lstinline|ICboot2|.

To do this, we consider different scenarios under the alternative hypothesis $H_1$  
consisting of time series from normal distributions with variance 1 with a change-point  
at time $\tau=20$ for mean differences ranging from 0.5 to 2.5 in steps of 0.25.

In each of these scenarios, we simulate \lstinline|Nsim2=1000| time series  
for each of which a change-point is indeed detected at the 5\% significance level,  
and for each of which two confidence intervals for the change-point estimate  
are determined using the functions \lstinline|ICboot1| and \lstinline|ICboot2|.

The proportion of cases where the time $\tau=20$ belongs to the interval obtained by \lstinline|ICboot1|  
(resp. \lstinline|ICboot2|)  
provides an estimate of the confidence level of such an interval.

The empirical mean of the lengths of the intervals obtained by \lstinline|ICboot1|  
(resp. \lstinline|ICboot2|)  
provides an estimate of the expected length of the intervals obtained by \lstinline|ICboot1|  
(resp. \lstinline|ICboot2|).

\blst
tau = 20  # change-point under H1
tabm1 = seq(0.5, 2.5, by = 0.25)  # differences in means before/after tau under H1
Nm1 = length(tabm1)

Nsim2 = 1000  # number of samples simulated under H1

coverage1 = numeric(Nm1)
avg_length1 = numeric(Nm1)
coverage2 = numeric(Nm1)
avg_length2 = numeric(Nm1)

for (j in 1:Nm1) {
  m1 = tabm1[j]
  # matrix XH1P of time series under H1
  # for which a change-point was detected at the 5
  XH1P = matrix(1:(Nsim2 * n), nrow = Nsim2)
  for (i in 1:Nsim2) {
    x = c(rnorm(tau), m1 + rnorm(n - tau))
    while (detect_change_point(x, loiVH0n)[1] > 0.05) {
      x = c(rnorm(tau), m1 + rnorm(n - tau))
    }
    XH1P[i, ] = x
  }
  
  # bootstrap confidence intervals using ICboot1 and ICboot2:
  clust = makeCluster(nclust)
  clusterExport(clust, c("XH1P", "ICboot1", "ICboot2",
                         "pvWMW", "pvWMWmat", "pvsWMW"))
  clusterEvalQ(clust, library(matrixTests)) 
  IC1 = parApply(clust, XH1P, 1, ICboot1)
  IC2 = parApply(clust, XH1P, 1, ICboot2)
  stopCluster(clust)
  
  coverage1[j] = sum(tau >= IC1[1, ] & tau <= IC1[2, ]) / Nsim2  # estimated coverage
  avg_length1[j] = sum(IC1[2, ] - IC1[1, ]) / Nsim2              # average CI length
  
  coverage2[j] = sum(tau >= IC2[1, ] & tau <= IC2[2, ]) / Nsim2  # estimated coverage
  avg_length2[j] = sum(IC2[2, ] - IC2[1, ]) / Nsim2              # average CI length
}
\end{lstlisting}

\begin{figure}[t]
    \centering
\includegraphics[scale=0.4]{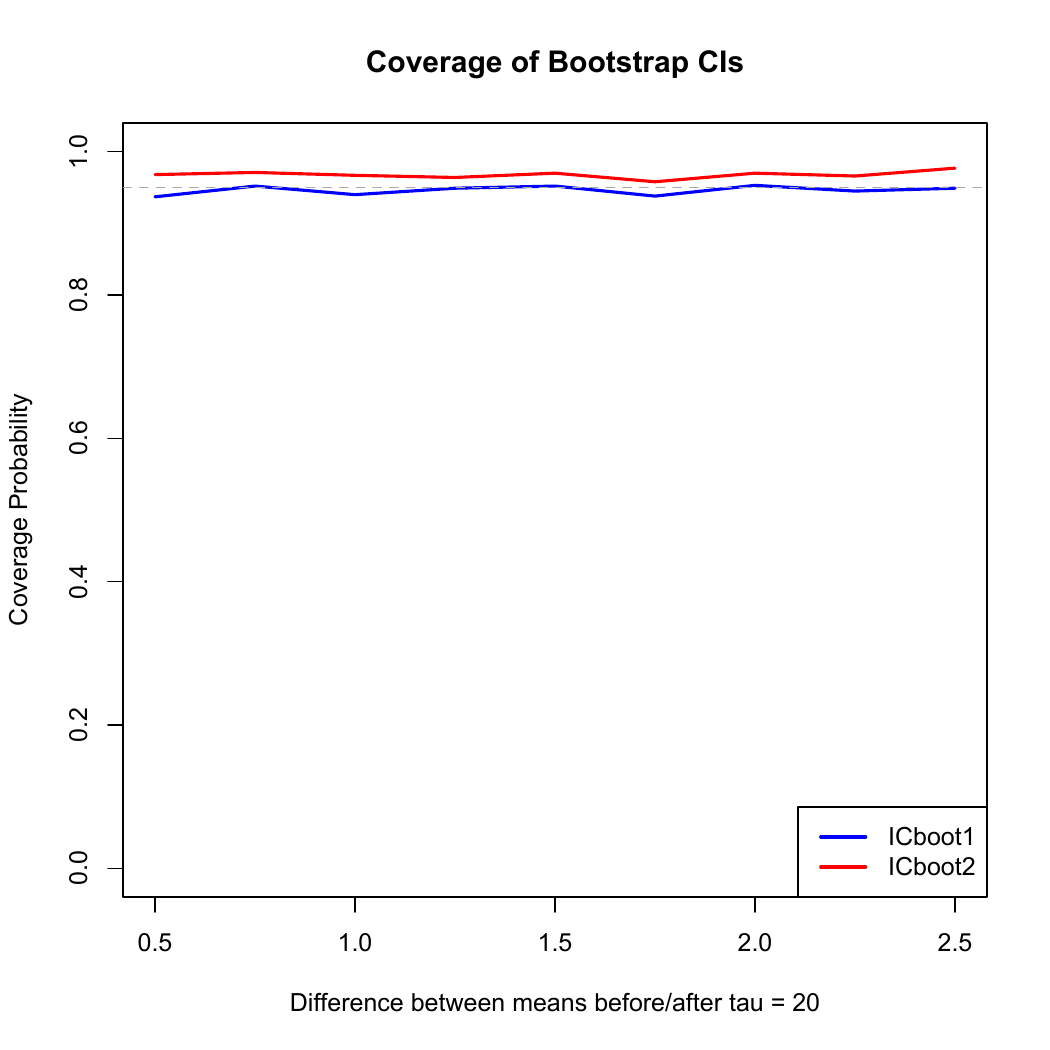}
\includegraphics[scale=0.4]{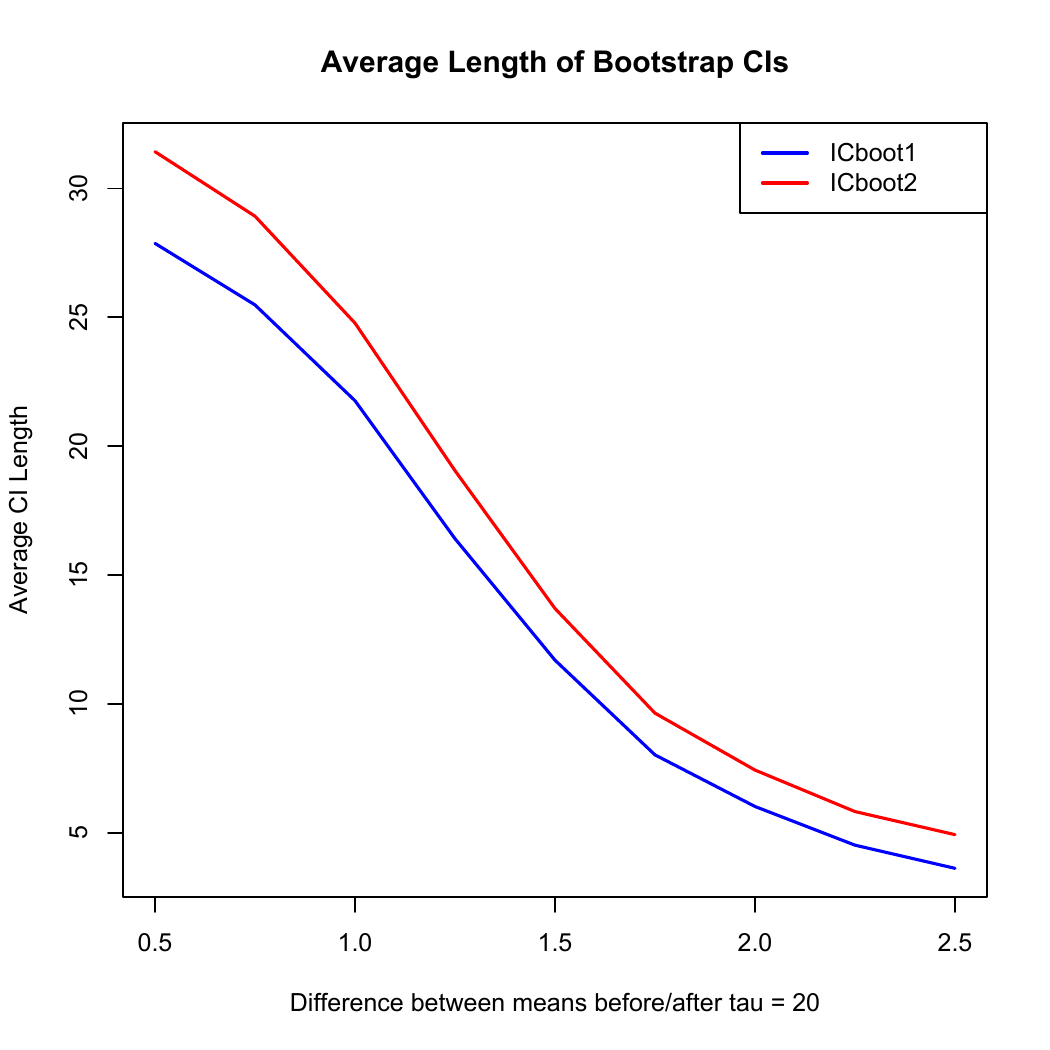}
  \caption{
Coverage probabilities and average lengths of the confidence intervals obtained with the R functions \lstinline|ICboot1| and \lstinline|ICboot2| as a function of the difference in mean $m_1$. Results obtained by simulating \lstinline|Nsim2=1000| time series with a change at time $\tau=20$.
}
    \label{IC20}
\end{figure}

Figure~\ref{IC20} shows the result obtained with $\tau=20$. As expected, the length of the confidence intervals decreases with $m_1$.
It is observed that the estimated confidence levels are close to $95\%$.  
With the function \lstinline|ICboot2|, there is a slight increase in these confidence levels,
but the average interval lengths become somewhat larger.

\begin{figure}[t]
    \centering
\includegraphics[scale=0.4]{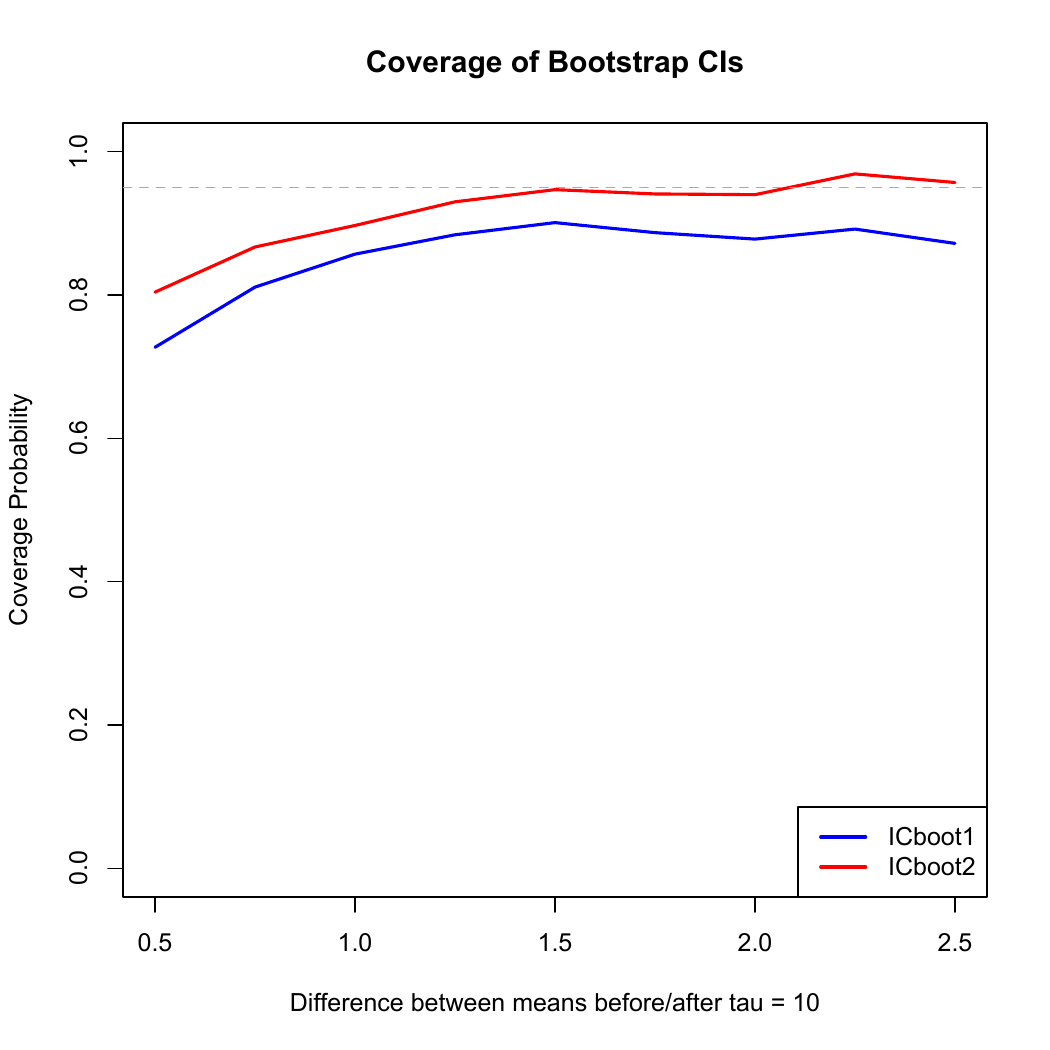}
\includegraphics[scale=0.4]{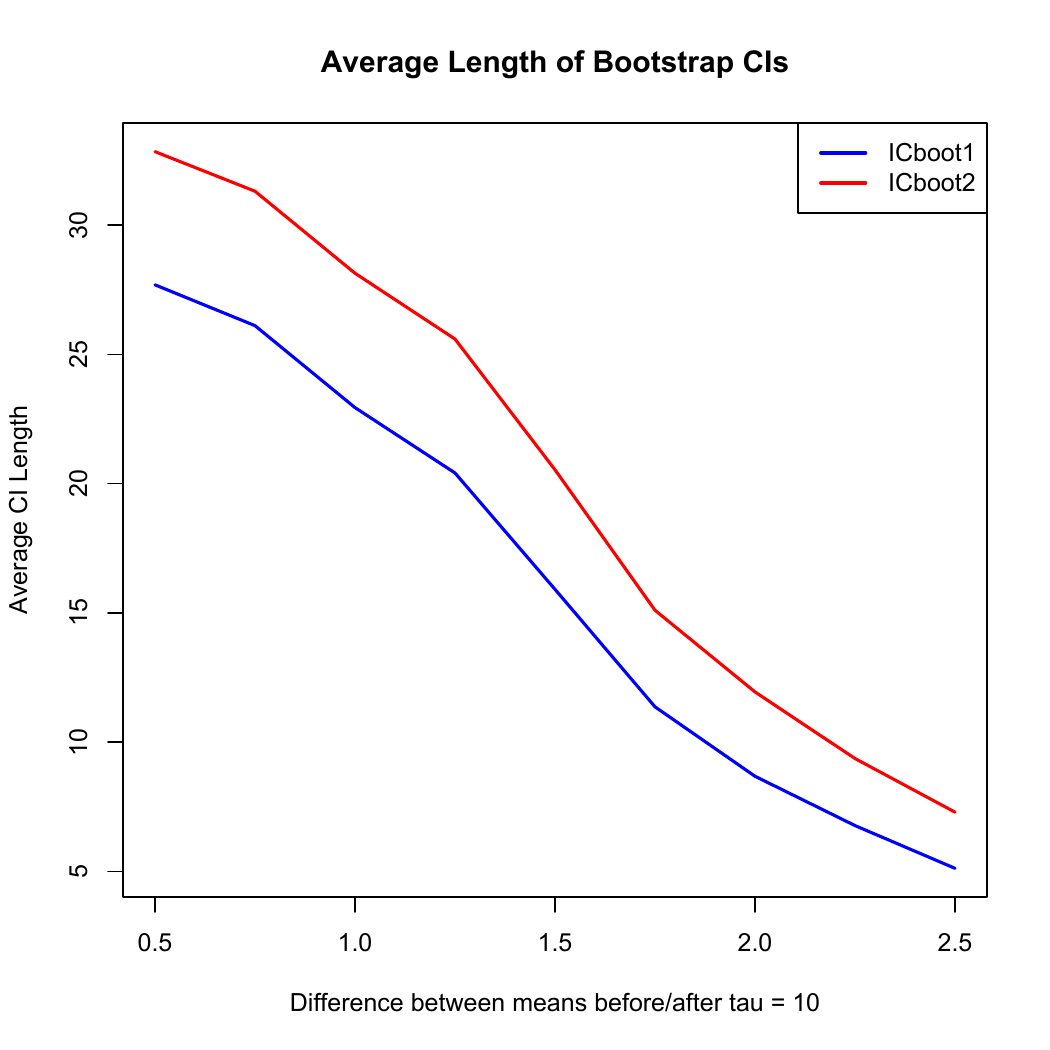}\\
\includegraphics[scale=0.4]{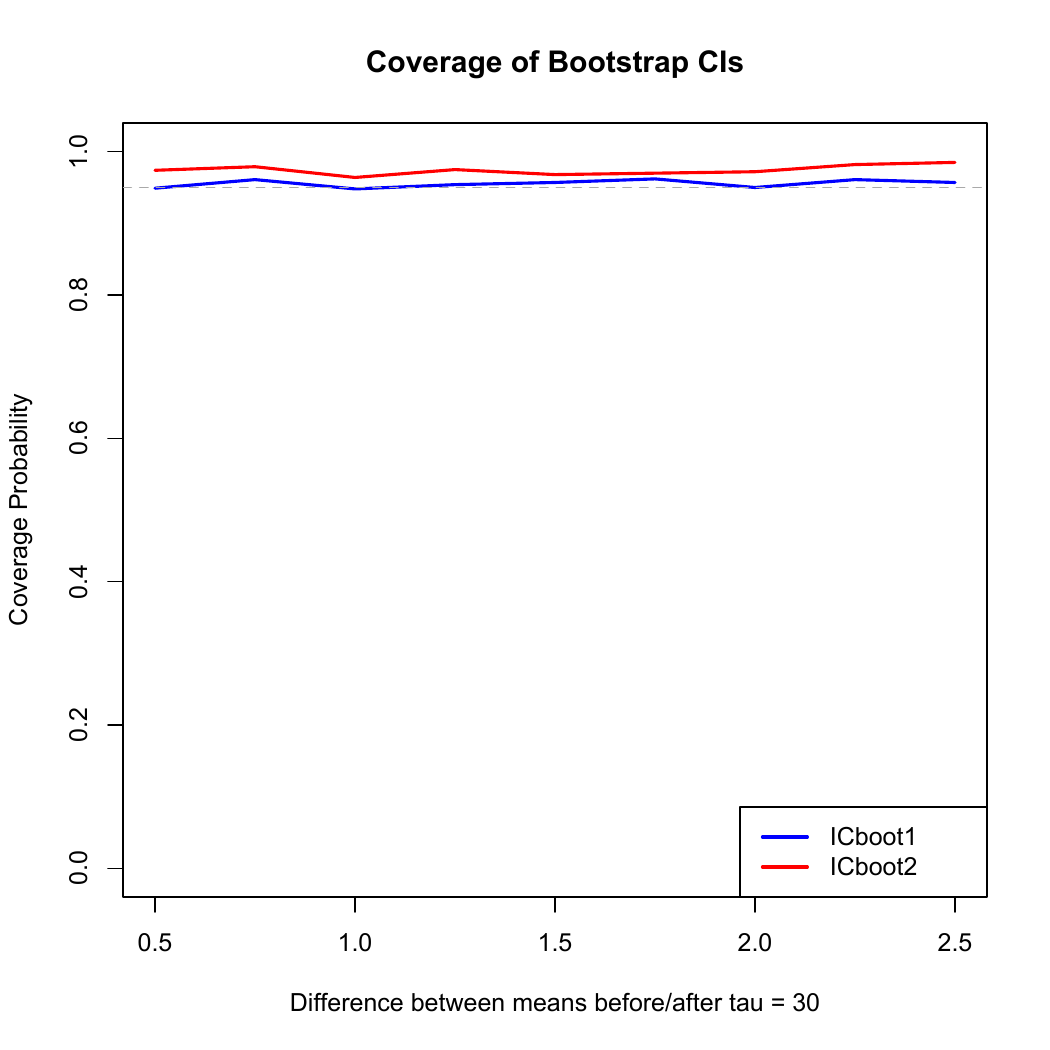}
\includegraphics[scale=0.4]{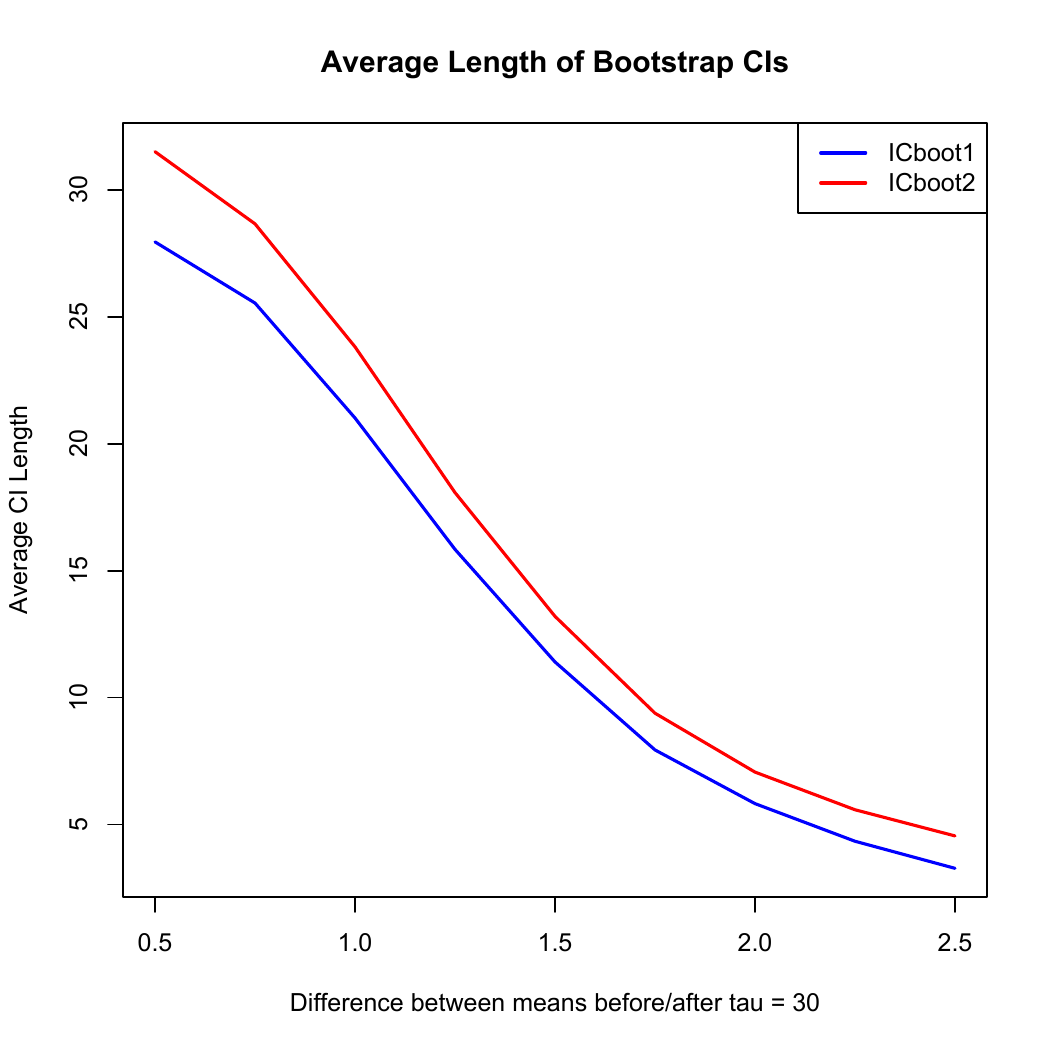}
  \caption{
Same as Figure~\ref{IC20} with $\tau=10$ (top) and $\tau=30$ (bottom).
}
    \label{IC10_30}
\end{figure}

\v5  
By choosing a change-point at time $\tau=10$, then at time $\tau=30$,  
four additional graphs are obtained (Figure~\ref{IC10_30}).
Both methods seems to provide approximatively $95\%$ confidence intervals when $\tau=30$, but the coverage probability is smaller than $95\%$ when $\tau=10$ and $m_1$ are small. Remark that it corresponds to challenging situations, with one sample with small size and a small difference between these two samples.

\v5
\n{\it Remark --}
In \cite{Valero}, confidence intervals are calculated using the function
\lstinline|ICboot1|.

\end{document}